\documentclass[twocolumn,showpacs,amsmath,amssymb,prd]{revtex4}
\usepackage{graphicx}
\newcommand{\beq}{\begin{equation}}
\newcommand{\eeq}{\end{equation}}
\newcommand{\beqn}{\begin{eqnarray}}
\newcommand{\eeqn}{\end{eqnarray}}
\newcommand{\tB}{\tilde{B}}

\newcommand{\sg}{\sqrt{\gamma}\,}
\newcommand{\ve}[1]{\mbox{\boldmath $#1$}}

\usepackage{graphicx}
\usepackage{dcolumn}
\usepackage{bm}
\usepackage{epsf}

\begin{document}

\newcommand{\geta}{\overline{\eta}}

\title{Relativistic Magnetohydrodynamics In Dynamical Spacetimes: \\ Numerical 
Methods And Tests}

\author{Matthew D. Duez}

\author{Yuk Tung Liu}

\author{Stuart L.\ Shapiro}
\altaffiliation{Also at Department of Astronomy \& NCSA, University of Illinois
at Urbana-Champaign, Urbana, IL 61801}

\author{Branson C. Stephens}

\affiliation{Department of Physics, University of Illinois at
Urbana-Champaign, Urbana, IL~61801}

\begin{abstract}
Many problems at the forefront of theoretical astrophysics require
the treatment of magnetized fluids in {\em dynamical}, strongly curved spacetimes. 
Such problems include the origin of gamma-ray bursts, magnetic
braking of differential rotation in nascent neutron stars arising
from stellar core collapse or binary neutron star merger, the formation of jets 
and magnetized disks
around newborn black holes, etc. To model these phenomena, all of
which involve both general relativity (GR) and magnetohydrodynamics (MHD),
we have developed a GRMHD code capable of evolving MHD fluids
in dynamical spacetimes. Our code solves the Einstein-Maxwell-MHD 
system of coupled equations in axisymmetry and in full 3+1 dimensions.  
We evolve the metric by integrating the BSSN
equations, and use a conservative, shock-capturing scheme to evolve
the MHD equations.  Our code gives accurate results in standard MHD code-test
problems, including magnetized shocks and magnetized Bondi flow.
To test our code's ability to evolve the MHD equations in a dynamical 
spacetime, we study the perturbations of a homogeneous, magnetized fluid
excited by a gravitational plane wave, and we find good agreement between the
analytic and numerical solutions.
\end{abstract}


\maketitle

\section{Introduction}
\label{intro}

Magnetic fields play a crucial role in determining the evolution of
many relativistic objects.  In any highly conducting astrophysical
plasma, a frozen-in magnetic field can be amplified appreciably by gas
compression or shear.  Even when an initial seed field is weak,
the field can grow in the course of time to significantly influence the gas
dynamical behavior of the system.  In problems where the self-gravity of
the gas can be ignored, simulations can be performed without numerically
evolving the spacetime metric.  Some accretion problems fall into this category.  In
many other problems, the effect of the magnetized fluid on the metric
cannot be ignored, and the two must be evolved self-consistently.
The final fate of many of these relativistic astrophysical systems,
and their distinguishing observational signatures, may hinge on the role
that magnetic fields play during the evolution. Some of these
systems are promising sources of gravitational radiation for detection
by laser interferometers such as LIGO, VIRGO, TAMA, GEO and LISA. 
Some also may be responsible for gamma-ray bursts. 
Examples of astrophysical scenarios involving
strong-field {\em dynamical} spacetimes in which MHD effects may play a
decisive role include the following:

$\bullet$ The merger of binary neutron stars. The coalescence can lead to
the formation of a {\it hypermassive} star supported by differential
rotation~\cite{bns_merger,stu03}.
While such a star may be dynamically stable against gravitational
collapse and bar formation, the radial stabilization due to
differential rotation is likely to be temporary. Magnetic braking and
viscosity combine to drive the star
to uniform rotation, even if the seed magnetic field and the viscosity
are small~\cite{mag_braking}. This process can lead to delayed
collapse and massive disk formation~\cite{dlss04}, accompanied
by a delayed gravitational wave burst.
MHD-related instabilities in differentially rotating stars might also drive
a gamma-ray burst~\cite{kr98}.

$\bullet$ Core collapse in a supernova. Core collapse may induce
differential rotation, even if the rotation of the progenitor at the
onset of collapse is only moderately rapid and almost uniform (see,
e.g.~\cite{ccollapse}, and references therein).
Differential rotation can wind up a frozen-in
magnetic field to high values, at which point it may provide a
significant source of stress, which could affect the explosion~\cite{mhd_sn}. 

$\bullet$ The generation of gamma-ray bursts (GRBs). Short-duration 
GRBs are thought to result from binary
neutron star mergers~\cite{narayan92}, or tidal disruptions of neutron stars
by black holes~\cite{ruffert99}, or hypergiant flares of `magnetars' 
associated with the soft gamma-ray repeaters~\cite{ngpf05}. 
Long-duration GRBs likely result from
the collapse of rotating, massive stars which form
black holes (`collapsars')~\cite{macfadyen99}.
In current scenarios, the burst is
powered by the extraction of rotational energy from the neutron star
or black hole, or from the remnant disk material formed around the black
hole~\cite{vlahakis01}. Strong magnetic fields provide a likely
mechanism for extracting this energy on the required timescale and
driving collimated GRB outflows in the form of relativistic
jets~\cite{meszaros97}. Even if the initial magnetic
fields are weak, they can be amplified to the required values by
differential motions~\cite{MRIs}.

$\bullet$ Supermassive black hole (SMBH) formation.  The origin
of the SMBHs observed in galaxies and quasars is one of the great
mysteries of contemporary astrophysics.  Several hypotheses for
the origin of SMBHs involve relativistic, self-gravitating fluids in which
magnetic fields can play an important role.  It is thought that
SMBHs start from smaller initial seed black holes, which grow to
supermassive size by a combination of accretion and mergers.  The seed
black holes might be provided by the collapse of massive
($M\sim 10^2M_{\odot}$) Population III stars~\cite{mr01}.  If so, magnetic
forces will affect the collapse leading to the formation of these seeds,
as well as their growth by accretion~\cite{gsm04,s05}.  Another
possibility is that
SMBHs form directly from the catastrophic collapse of supermassive stars
(SMSs)~\cite{r84}.  This collapse will proceed differently, depending on
whether the SMS rotates uniformly or differentially~\cite{ns01}.
Magnetic fields and turbulence provide the
principle mechanisms that can damp differential rotation in such
stars~\cite{zn71} and thus determine their ultimate fate.

$\bullet$ The r-mode instability in rotating neutron stars. This
instability has been proposed as a possible mechanism for
limiting the angular velocities in neutron stars and producing
observable quasi-periodic gravitational waves~\cite{r_modes}.  However,
preliminary calculations suggest that if
the stellar magnetic field is strong enough, r-mode oscillations will
not occur~\cite{rlms01}. Even if the initial field is weak,
fluid motions produced
by these oscillations may amplify the magnetic field and eventually
distort or suppress the r-modes altogether. (R-modes may also be
suppressed by non-linear mode coupling~\cite{saftw02} or
hyperon bulk viscosity~\cite{jl02}.)

$\bullet$ Massive disk accretion.  The importance of magnetic effects
on gas accretion onto a black hole is well known.  In many cases,
the density of the accreting material is small enough that its
effect on the spacetime geometry is negligible.  Such systems can be
studied by evolving the gas on the stationary 
Kerr spacetime background produced by the central
black hole.  There are, however, situations in which accretion disks
with masses comparable to that of the central black hole can be formed.  Examples
include the collapse of rapidly rotating stars or
supermassive stars~\cite{ss02},
and neutron star merger (especially when the two neutron stars have unequal
masses~\cite{stu03}). In these cases, the spacetime is dynamical and 
Einstein's equations for the metric must be evolved along with the 
MHD equations.

In the recent years, numerical codes which evolve the general relativistic
MHD equations on fixed Schwarzschild or Kerr black hole spacetimes have
been developed by Yokosawa~\cite{y93}, Koide {\it et al}~\cite{ksk99}, Komissarov~\cite{k04},
De Villiers and Hawley~\cite{dVh03}, and Gammie {\it et al}~\cite{HARM}.  These
codes have been used to study the structure of accretion flows onto Kerr
black holes~\cite{mg04,dVhk03}, the Blandford-Znajek effect in low-density
regions near the hole~\cite{k04,k05}, and the formation of GRB jets~\cite{myks04}.

In contrast to the above effort, few attempts have been made to simulate
relativistic MHD flows in dynamical spacetimes.  One major attempt
was by Wilson, thirty years ago~\cite{w75}.  He simulated the collapse of a
rotating SMS with a frozen-in poloidal magnetic field in axisymmetry 
using a code which assumed the conformal flatness approximation for the 
spatial metric. 
No gravitational radiation is allowed in this approximation.
Wilson's work was generalized by Nakamura, Oohara and Kojima in 
1987~\cite{nok87}. They studied the effect of poloidal  magnetic 
fields on the collapse of nonrotating SMSs in full GR in axisymmetry. 
Since the stars are nonrotating, toroidal fields are not 
generated, which simplifies the calculation. 
Since then, no other simulations of this kind have been attempted.  
However, in anticipation of future
numerical work, formulations of the coupled Einstein-Maxwell-MHD equations
were proposed by Sloan and Smarr~\cite{ss85}, by Zhang~\cite{z89}, and by
Baumgarte and Shapiro~\cite{bs03}.

In this paper, we present the first code capable of evolving the
Einstein-Maxwell-MHD equations without approximation in both two dimensions (axisymmetry) and three
dimensions.  Our code is based on the Baumgarte-Shapiro-Shibata-Nakamura
(BSSN) formulation of the 3+1 Einstein field equations~\cite{BSSN}.  In
previous papers, we have evolved the BSSN equations coupled
to a perfect fluid~\cite{dmsb03}, and we have applied our code to simulate
stellar collapse and binary neutron star inspiral~\cite{dmsb03,mdsb04}.
We then
generalized our code to study fluids with shear viscosity~\cite{dlss04},
and we implemented black hole excision techniques to study the collapse
of fluid stars to black holes~\cite{dsy04}.  In this paper, we have
completely reformulated the hydrodynamics sector of our code in order
to improve its accuracy and shock-handling capability.  We have 
extended the code to allow for a magnetic field frozen into a 
perfectly conducting fluid in the MHD approximation.

In Sec.~II, we present the evolution equations integrated by our code.
In Sec.~III, we discuss the adopted numerical techniques. We
perform a variety of code tests in Sec.~IV, including magnetized shocks, 
a magnetized
Bondi accretion flow, and a linear gravitational wave that 
excites MHD waves.

\section{Formalism}

\subsection{Evolution of the gravitational fields}

Throughout this paper, Latin indices denote spatial components
(1-3) and Greek indices denote spacetime components (0-3). We write
the metric in the form
\begin{equation}
ds^2 = -\alpha^2 dt^2 + \gamma_{ij}(dx^i+\beta^idt)(dx^j+\beta^jdt),
\end{equation}
where $\alpha$, $\beta^i$, and $\gamma_{ij}$ are the lapse, shift, and
spatial metric, respectively.  The extrinsic curvature $K_{ij}$ is
defined by
\begin{equation}
\label{Kij}
(\partial_t - {\mathcal{L}}_{\beta})\gamma_{ij} = -2\alpha K_{ij},
\end{equation}
where ${\mathcal{L}}_{\beta}$ is the Lie derivative with respect to
$\beta^i$.  We adopt geometrized units, so that $G = c = 1$.
We evolve $\gamma_{ij}$ and $K_{ij}$
using the BSSN formulation~\cite{BSSN}.  The fundamental variables for
BSSN evolution are
\begin{eqnarray}
  \phi &\equiv& {1\over 12}\ln[\det(\gamma_{ij})]\ , \\
  \tilde\gamma_{ij} &\equiv& e^{-4\phi}\gamma_{ij}\ , \\
  K &\equiv& \gamma^{ij}K_{ij}\ , \\
  \tilde A_{ij} &\equiv& e^{-4\phi}(K_{ij} - {1\over 3}\gamma_{ij}K)\ , \\
  \tilde\Gamma^i &\equiv& -\tilde\gamma^{ij}{}_{,j}\ .
\end{eqnarray}
The evolution and constraint equations for
these fields are summarized in~\cite{BSSN,dsy04}.  In the
presence of mass-energy, these evolution equations
contain the following source terms:
\begin{eqnarray}
  \rho &=& n_{\alpha}n_{\beta}T^{\alpha\beta}\ , \nonumber \\
  S_i  &=& -\gamma_{i\alpha}n_{\beta}T^{\alpha\beta}\ , \label{source_def} \\
  S_{ij} &=& \gamma_{i\alpha}\gamma_{j\beta}T^{\alpha\beta}\ ,\nonumber
\end{eqnarray}
where $T^{\alpha\beta}$ is the stress tensor, and
$n^{\alpha} = (\alpha^{-1},-\alpha^{-1} \beta^i)$ is the time-like
unit vector normal to the $t=$ constant time slices.  
Note that, since the electromagnetic fields contribute
to $T^{\alpha\beta}$, they will contribute to $\rho$, $S_i$, and
$S_{ij}$ as shown in Eq.~(\ref{source_def}).

In order to evolve the 3+1 Einstein equations forward in time, one
must choose lapse $\alpha$ and shift $\beta^i$ functions, which
specify how the spacetime is foliated. 
The lapse and shift must be chosen in such a way that the total
system of evolution equations is stable.  It is also desirable that
the adopted gauge conditions make the evolution appear as stationary
as possible.  As in~\cite{dsy04}, we use the following hyperbolic
driver conditions:
\begin{eqnarray}
\label{hb_lapse_nok3}
\nonumber
\partial_t \alpha &=& \alpha {\cal A} \\
\partial_t {\cal A} &=& -a_1(\alpha\partial_tK \\
\nonumber
   & &\ + a_2\partial_t\alpha + a_3 e^{-4\phi}\alpha K)\ . \\
\label{hb_shift}
\partial^2_t\beta^i &=& b_1(\alpha\partial_t\tilde\Gamma^i
 - b_2\partial_t\beta^i)\ ,
\end{eqnarray}
where $a_1$, $a_2$, $a_3$, $b_1$, and $b_2$ are freely specifiable
constants.  We usually choose $a_1 = b_1 = 0.75$, $a_2 = b_2 = 0.34M^{-1}$,
$a_3 = 1$.  (There are exceptions.  For example, when evolving a collapsing
system, it is better to use a smaller $b_1$.  
This prevents ``blowing out'' of the coordinate system, a well-known
effect~\cite{sbs00,dmsb03} which can spoil grid resolution in the center of the
collapsing object.)

After the initial time, we do not enforce the constraint equations, but
rather monitor them as a check on the accuracy of our evolution of the
metric.  Another check on the metric evolution is the conservation of the
ADM mass $M$ and angular momentum $J$ of the spacetime, accounting 
for losses due to gravitational radiation or rest-mass outflow 
from the grid (both are negligible for the tests reported here).  
The formulae for $M$ and $J$ are given in~\cite{dmsb03} in terms of 
volume integrals over the whole space. 
For spacetimes containing black holes in which excision is employed, we 
use the formulae given in~\cite{dsy04}, which consist of a surface 
integral over a sphere enclosing the excision region, plus a volume 
integral from the surface to spatial infinity. 

\subsection{Evolution of the electromagnetic fields}

The electromagnetic stress-energy tensor $T^{\mu\nu}_{\rm em}$ is given 
by 
\beq
 T^{\mu\nu}_{\rm em} = \frac{1}{4\pi} \left( 
F^{\mu\lambda}F^{\nu}{}_{\lambda} - \frac{1}{4}
g^{\mu\nu}F_{\alpha\beta}F^{\alpha\beta} \right) \ .
\eeq
We decompose the Faraday tensor $F^{\mu\nu}$ as
\beq
  F^{\mu \nu} = n^{\mu} E^{\nu} - n^{\nu} E^{\mu} + n_{\gamma} 
\epsilon^{\gamma \mu \nu \delta} B_{\delta} \ ,
\label{eq:FabEB}
\eeq
so that $E^{\mu}$ and $B^{\mu}$ are the electric and magnetic
fields measured by a normal observer $n^{\mu}$.  Both fields
are purely spatial ($E^{\mu}n_{\mu} = B^{\mu}n_{\mu} = 0$), and
one can easily show that
\beqn
  E^{\mu} = F^{\mu \nu} n_{\nu} \ \ \ , \ \ \ 
  B^{\mu} = \frac{1}{2} \epsilon^{\mu \nu \kappa \lambda} n_{\nu} 
F_{\lambda \kappa} = n_{\nu} F^{* \nu \mu} \ ,
\label{eq:EB}
\eeqn
where 
\beq
   F^{* \mu \nu} = \frac{1}{2} \epsilon^{\mu \nu \kappa \lambda} 
F_{\kappa \lambda} 
\eeq
is the dual of $F^{\mu \nu}$.
In terms of $E^{\mu}$ and $B^{\mu}$, the electromagnetic stress tensor
is given by
\beqn
  T^{\mu\nu}_{\rm em} &=& \frac{1}{8\pi}(g^{\mu\nu} + 2 n^{\mu}n^{\nu})
(E^{\lambda}E_{\lambda}+B^{\lambda}B_{\lambda}) \cr
& & -\frac{1}{4\pi} \left( B^{\mu}B^{\nu}+E^{\mu}E^{\nu} \right) \cr \cr
& & + \frac{1}{4\pi} n^{(\mu}\epsilon^{\nu)\sigma\delta}
E_{\sigma}B_{\delta} \ . \label{eq:Tabem}
\eeqn
Along with the electromagnetic field, we also assume the presence of
a perfect fluid with rest density $\rho_0$, pressure $P$, and 4-velecity
$u^{\mu}$, so that the total stress-energy tensor is
\beq
  T^{\mu \nu} = \rho_0 h u^{\mu} u^{\nu} + P g^{\mu \nu}
  + T_{\rm em}^{\mu \nu}\ ,
\eeq
where the specific enthalpy $h$ is related to the specific internal energy
$\epsilon$ by $h = 1 + \epsilon + P/\rho_0$.

For most applications of interest in relativistic astrophysics, we can assume
perfect conductivity.  In the limit of infinite conductivity, Ohm's law yields
the MHD condition:
\beq
u_{\mu}F^{\mu\nu} = 0\ . \label{cond:MHD}
\eeq
The electric and magnetic fields measured by an observer comoving 
with the fluid are [cf.\ Eq.~(\ref{eq:EB})] 
\beqn
  E^{\mu}_{(u)} = F^{\mu \nu} u_{\nu}  \ \ \ , \ \ \ 
  B^{\mu}_{(u)} = u_{\nu} F^{* \nu \mu} \ .
\label{eq:EuBu}
\eeqn
The ideal MHD condition~(\ref{cond:MHD}) is equivalent to the 
statement that the electric field observed in the fluid's rest 
frame vanishes ($E^{\mu}_{(u)}=0$). 
Note that $B^{\mu}_{(u)}$ is orthogonal to $u_{\mu}$, 
i.e.\ $u_{\mu} B^{\mu}_{(u)}=0$.
We can express $F^{\mu\nu}$ in terms of $B^{\mu}_{(u)}$ as  
[cf.\ Eq.~(\ref{eq:FabEB})]
\beq
  F^{\mu \nu} = u_{\gamma}
\epsilon^{\gamma \mu \nu \delta} B^{(u)}_{\delta} \ .
\label{eq:FabBu}
\eeq
Taking the dual of Eq.~(\ref{eq:FabBu}), we obtain 
\beq
  F^{* \mu \nu} = B^{\mu}_{(u)} u^{\nu} - B^{\nu}_{(u)} u^{\mu} \ . 
\label{eq:FstarBu}
\eeq
We define the projection operator $P_{\mu \nu} = g_{\mu \nu} + u_{\mu} u_{\nu}$.  
Since $B^{\mu}_{(u)}$ is orthogonal to $u_{\mu}$, we have 
$P^{\mu}{}_{\nu} B^{\nu}_{(u)}=B^{\mu}_{(u)}$. It follows from 
Eqs.~(\ref{eq:EB}) and~(\ref{eq:FstarBu}) that 
\beq
  P^{\mu}{}_{\nu} B^{\nu} = P^{\mu}{}_{\nu} n_{\lambda} 
(B^{\lambda}_{(u)} u^{\nu} - B^{\nu}_{(u)} u^{\lambda}) 
 = -n_{\lambda} u^{\lambda} B^{\mu}_{(u)} \ .
\eeq
Hence we have 
\beq
  B^{\mu}_{(u)} = -\frac{P^{\mu}{}_{\nu} B^{\nu}}{n_{\nu} u^{\nu}} \ .
\label{bintermsofB}
\eeq
Evaluating the time and spatial components of Eq.~(\ref{bintermsofB}) 
gives
\beqn
  B^0_{(u)} &=& u_i B^{i}/\alpha \ , \\
  B^i_{(u)} &=& \frac{B^i/\alpha + B^0_{(u)} u^i}{u^0} \ . 
\label{biintermsofBi}
\eeqn

The evolution equation for the magnetic field can be obtained in
conservative form by taking the dual of Maxwell's equation
$F_{[\mu \nu,\lambda]}=0$.  One finds
\beq
  \nabla_{\nu} F^{* \mu \nu} = \frac{1}{\sqrt{-g}} \partial_{\nu} 
(\sqrt{-g}\, F^{*\mu \nu}) = 0 \ , \label{Maxwell}
\eeq
where $\sqrt{-g} = \alpha \sg$.
Note that $F^{*i0} = B^i/\alpha$ [see Eq.~(\ref{eq:EB})]. 
The time component of
Eq.~(\ref{Maxwell}) gives the no-monopole constraint 
\beq
  \partial_j \tilde{B}^j =0 \ , 
\label{divB}
\eeq
where 
\beq
  \tilde{B}^j = \sqrt{\gamma}\, B^j \ .
\eeq
The spatial components of Eq.~(\ref{Maxwell}) give the induction equation 
\beq
  \partial_t \tilde{B}^i + \partial_j [\sqrt{-g} (u^j B^i_{(u)} - u^i 
B^j_{(u)})]=0 \ .
\eeq 
It follows from Eq.~(\ref{biintermsofBi}) that 
\beq
  u^j B^i_{(u)} - u^i B^j_{(u)} = (v^j B^i - v^i B^j)/\alpha \ , 
\eeq
where $v^i = u^i/u^0$. Hence the induction equation can be written as 
\beq
  \partial_t \tilde{B}^i + \partial_j (v^j \tilde{B}^i - v^i 
\tilde{B}^j)=0 \ .
\eeq 

\subsection{Evolution of the hydrodynamics fields}

In the literature, a magnetic 4-vector $b^{\mu}$ is often introduced. 
It is related to $B^{\mu}_{(u)}$ by 
\beq
  b^{\mu} = \frac{B^{\mu}_{(u)}}{\sqrt{4\pi}} \ .
\label{eq:baBa}
\eeq
In the MHD limit, $T^{\mu \nu}_{\rm em}$ can be expressed simply in terms of
$b^{\mu}$ as [cf.\ Eq.~(\ref{eq:Tabem})]
\beq
  T^{\mu \nu}_{\rm em} = b^2 u^{\mu} u^{\nu} + \frac{1}{2} b^2 g^{\mu \nu} 
 - b^{\mu} b^{\nu} \ , 
\eeq
and the total stress tensor is given by
\beq
  T^{\mu \nu} = (\rho_0 h +b^2) u^{\mu} u^{\nu} + \left( P + \frac{b^2}{2} 
\right) g^{\mu \nu} - b^{\mu} b^{\nu} \ .
\eeq
The evolution equations for the fluid are given by the baryon number
conservation equation $\nabla_{\nu} (\rho_0 u^{\nu})=0$
and the energy-momentum conservation equation
$\nabla_{\nu} T_{\mu}{}^{\nu}=0$.  Conservation of baryon number gives
\beq
  \partial_t \rho_* + \partial_j (\rho_* v^j) = 0 \ ,
\eeq
where $\rho_* = \alpha \sg \rho_0 u^0$.
The spatial components of the energy-momentum conservation 
equation give the momentum equation
\beq
  \partial_t \tilde{S}_i 
 + \partial_j (\alpha \sqrt{\gamma}\, T^j{}_i) = \frac{1}{2} \alpha \sqrt{\gamma} 
\, T^{\alpha \beta} g_{\alpha \beta,i} \ ,
\eeq
where we have defined the momentum density variable 
\beqn
\label{momdef}
  \tilde{S}_i &=& \sqrt{\gamma} S_i = \alpha \sg T^0{}_i  \cr 
  &=& (\rho_* h + \alpha u^0 \sg b^2) u_i - \alpha \sg b^0 b_i \ .
\eeqn

The time component of the energy-momentum conservation 
equation gives the energy equation. 
Following Font {\it et al}~\cite{fmst00}, we use the energy variable 
\beqn
  \tilde{\tau} = \sqrt{\gamma}\, n_{\mu} n_{\nu} T^{\mu \nu} - \rho_*
 = \alpha^2 \sqrt{\gamma}\, T^{00} - \rho_* 
\label{taudef}
\eeqn
The energy equation is then given by
\beq
  \partial_t \tilde{\tau} + \partial_i ( \alpha^2 \sqrt{\gamma}\, T^{0i} 
-\rho_* v^i) = s \ ,
\label{fonts_eng_eq}
\eeq
where the source term $s$ is
\beqn
  s &=& -\alpha \sqrt{\gamma}\, T^{\mu \nu} \nabla_{\nu} n_{\mu}  \cr
   &=& \alpha \sqrt{\gamma}\, [ (T^{00}\beta^i \beta^j + 2 T^{0i} \beta^j 
+ T^{ij}) K_{ij} \cr
 & & - (T^{00} \beta^i + T^{0i}) \partial_i \alpha ]\ .
\eeqn

To complete the system of equations, it remains only to specify the
equation of state (EOS) of the fluid.  In this paper, we adopt 
a $\Gamma$-law EOS
\begin{equation}
\label{ideal_P}
P = (\Gamma - 1)\rho_0\epsilon,
\end{equation}
where $\Gamma$ is a constant.
We choose the $\Gamma$-law EOS because it simplifies some of the 
calculations, it is applicable to many cases of interest, and it is 
a standard choice for demonstrating new computational techniques
in the numerical relativity literature. We note that all equations 
in this section apply for any equation of state. Generalization 
to a more realistic EOS is not difficult, and we plan to use more 
realistic EOSs in some of our future work.

To summarize, the evolution equations for the magnetohydrodynamic
variables are
\vskip 0.3cm
\fbox{
\parbox{8cm}{
\beq
  \partial_t \rho_* + \partial_j (\rho_* v^j) = 0 \ ,
\label{eq:summ1}
\eeq
\beq
  \partial_t \tilde{\tau} + \partial_i ( \alpha^2 \sqrt{\gamma}\, T^{0i} 
-\rho_* v^i) = s \ ,
\eeq
\beq
  \partial_t \tilde{S}_i 
 + \partial_j (\alpha \sqrt{\gamma}\, T^j{}_i) = \frac{1}{2} \alpha \sqrt{\gamma} 
\, T^{\alpha \beta} g_{\alpha \beta,i} \ ,
\eeq
\beq
  \partial_t \tilde{B}^i + \partial_j (v^j \tilde{B}^i - v^i \tilde{B}^j)=0 \ .
\label{eq:summ4}
\eeq
}}

\section{Implementation}

We use a cell-centered Cartesian grid in our three-dimensional simulations. 
Sometimes, symmetries can be invoked to reduce the integration domain. For
octant symmetric systems, we evolve only the upper octant; for
equatorially symmetric systems, we evolve only the upper half-plane.  For
axisymmetric systems, we evolve only the $x-z$ plane [a (2+1)D problem]. 
In axisymmetric evolutions, 
we adopt the Cartoon method~\cite{cartoon} for evolving the BSSN equations, 
and use a cylindrical grid for evolving the induction and MHD equations.

Our technique for evolving the metric fields is described in our
earlier papers~\cite{dmsb03,dsy04,dlss04}, so we focus here on
our MHD algorithms.
The goal of this part of the numerical evolution is to determine the
fundamental MHD variables ${\bf P} = (\rho_0,P,v^i,B^i)$,
called the {\it ``primitive'' variables}, at future times, given
initial values of ${\bf P}$. 
The evolution equations (\ref{eq:summ1})--(\ref{eq:summ4}) are 
written in conservative form,
i.e. they give the time derivatives of the {\it ``conserved'' variables}
${\bf U}({\bf P}) = (\rho_{\star},\tilde\tau,\tilde{S}_i,\tB^i)$ in terms of
{\it source variables} ${\bf S}({\bf P})$ and the divergence of
{\it flux variables} ${\bf F}({\bf P})$: 
\beq
\partial_t {\bf U} + \nabla\cdot{\bf F} = {\bf S}\ ,
\label{cons}
\eeq
where ${\bf F}({\bf P})$ and ${\bf S}({\bf P})$ are not explicit
functions of derivatives of the primitive variables, although
they are explicit functions of the metric and its derivatives.

There are several ways of evolving this system.  Conservative
schemes evolve ${\bf U}$ with the equations (\ref{cons}).  The advantage
of this is that highly accurate shock-capturing methods can be applied to
this set of equations.  The disadvantage is that, after each timestep,
one must recover ${\bf P}$ by numerically solving the system of
equations ${\bf U} = {\bf U}({\bf P})$, which can be complicated and
computationally expensive.  Non-conservative schemes, on the other
hand, evolve variables which are more simply related to ${\bf P}$ but
whose evolution equations are not of the form of Eq.~(\ref{cons}).  In such
schemes, high-resolution shock-capturing methods cannot be used and
artificial viscosity must be introduced for handling discontinuities,
but the recovery of ${\bf P}$ is fairly straightforward.  After
implementing both types of schemes, we found our conservative scheme
to be more stable when strong magnetic fields are present.
We evolve Eq.~(\ref{cons}) using a three-step iterated Crank-Nicholson scheme 
as in several of our previous papers~\cite{dmsb03}.  This scheme is 
second order in time and will be stable if 
$\Delta t < \min(\Delta x^i)/c_{\rm max}$,
where in our case $c_{\rm max}$ is the speed of light.  In most
applications below, we set $\Delta t = 0.5 \min(\Delta x^i)$.  For each Crank-Nicholson
substep, we first update the gravitational field variables (the BSSN 
variables).  We then update the electromagnetic fields $B^i$ by integrating the induction
equation.  Next, the remaining MHD variables ($\rho_{\star}$, $\tilde{\tau}$, and
$\tilde{S}_i$) are updated.  Finally, we use these updated values to reconstruct
the primitive variables on the new timestep.

\subsection{The reconstruction step}
We have implemented an approximate Riemann solver to handle the flux term
in Eq.~(\ref{cons}).  Below, we will demonstrate how the flux $f$ is 
calculated for a given conserved variable, $u$.  For simplicity, we will
consider the one-dimensional case.  The generalization to three dimensions is
straightforward.  The first step in 
calculating this flux is to compute ${\bf P}_L = {\bf P}_{i+1/2-\epsilon}$ and
${\bf P}_R = {\bf P}_{i+1/2+\epsilon}$, i.e. the primitive variables
to the left and right of the grid cell interface.  We have implemented several
methods for computing ${\bf P}_L$ and ${\bf P}_R$.

1) {\it Monotonized central (MC) reconstruction} \\
This method~\cite{vL77} gives second-order accurate results at most points.  For a
given primitive variable $p$, one sets
\beqn
p_L &=& p_i + \nabla p_i \Delta x/2 \nonumber \\
p_R &=& p_{i+1} - \nabla p_{i+1} \Delta x/2\ .
\eeqn
Here, $\nabla p$ is the slope-limited gradient of $p$: 
\mbox{$\nabla p = \Delta x^{-1}{\rm MC}(p_{i+1}-p_i,p_i-p_{i-1})$}, where
\beq
\label{eq:mc}
{\rm MC}(a,b) = \left\{ \begin{array}{ll}
                     0 & \mbox{if $ab\leq 0$\ ,} \\
		     {\rm sign}(a)\min(2|a|,2|b|,|a+b|/2) & \mbox{otherwise\ .}
		     \end{array}
          \right.
\eeq
This scheme becomes first-order accurate at extrema of $p$.

2) {\it Convex essentially non-oscillatory (CENO) reconstruction} \\
In this scheme~\cite{lo98}, one uses polynomial (usually quadratic) interpolation
to find cell face values.  For smooth monotonic functions, these values are
accurate to third order in $\Delta x$.  As in the above method, $p_L$ and $p_R$
usually differ, and the scheme becomes first order at extrema of $p$. 
See~\cite{dZb02,dZbl03} for details of this reconstruction method.

3) {\it Piecewise Parabolic (PPM) reconstruction} \\
For smooth monotonic functions on uniform grids,
PPM reconstructs face values to
third-order accuracy in $\Delta x$.  In this scheme, $p_L$ and $p_R$ are
equal except in special circumstances, usually involving shocks~\cite{PPM}.
We have made slight modifications to this scheme, the details of
which are discussed in Appendix~\ref{ppm_appendix}.  These changes
do not affect the fluid evolution in any of the applications
below except that of unmagnetized stars.  In this particular case, it is necessary
to distinguish between the standard PPM scheme as given in~\cite{PPM}
and ours, and so we refer to the former as PPM and the latter as
PPM$^+$.

We note that, even when using higher order reconstruction schemes such
as CENO and PPM, our overall evolution scheme remains second-order
accurate in space and time.  This is because our finite differencing of Eq.~(\ref{cons})
is only second-order accurate (although this could be improved), and
the BSSN variables are only evolved with second-order accuracy. 
Nevertheless, higher order reconstruction schemes can provide more accurate
results for some applications.

\subsection{The Riemann solver step}
Next, we take the reconstructed data as initial data for a
piecewise constant Riemann problem, with ${\bf P} = {\bf P}_L$ on
the left of the interface, and ${\bf P} = {\bf P}_R$ on
the right of the interface.  The net flux at the cell interface
is given by the solution to this Riemann problem.

We use the HLL (Harten, Lax, and van Leer) approximate Riemann 
solver~\cite{HLL}.  The HLL solver is
one of the simplest shock-capturing schemes as it does not
require knowledge of the eigenvectors of the system.  Nevertheless, when 
coupled to a higher order reconstruction method such as PPM, even simpler
Riemann solvers, let alone HLL, have
been shown to perform with an accuracy comparable to
more sophisticated solvers in shock tube problems~\cite{dZb02,lfim04} and
in binary neutron star simulations~\cite{s05pc}.
To compute the HLL fluxes,
one only needs to
provide a maximum left-going wave speed $c_+$ and a maximum right-going
wave speed $c_-$ on both sides of the interface.  Defining
$c_{\rm max} \equiv \max(0,c_{+R},c_{+L})$ and
$c_{\rm min} \equiv -\min(0,c_{-R},c_{-L})$, the HLL flux is given by
\beq
\label{eq:hll}
f_{i+1/2} = {c_{\rm min} f_R + c_{\rm max} f_L
  - c_{\rm min}c_{\rm max} (u_R - u_L) \over 
  c_{\rm max} + c_{\rm min} }\ .
\eeq

We compute the wave speeds $c_{\pm}$ as described in Section 3.2 of~\cite{HARM}.  Only the 
maximum wave speeds in either direction along the three
coordinate axes are required.  To determine the speeds in the $x$ direction, one
solves the dispersion relation for MHD waves with wave vectors of the form
\beq
\label{kform}
k_{\mu} = (-\omega,k_1,0,0).
\eeq
The wave speed is simply the phase
speed $\omega/k_1$.  The speeds along $y$ and $z$ are computed in a
similar way.  As in ~\cite{HARM}, we replace the full dispersion relation
by a simpler expression which overestimates the maximum speeds by a
factor of $\le 2$ (thus making the evolution stabler, but also more
diffusive).  In the frame comoving with the fluid, the approximate
dispersion relation for MHD waves is
\beq
\label{eq:dispersion}
\omega_{\rm cm}^2 = \left[v_A^2 + c_s^2\left(1-v_A^2\right)\right]k_{\rm cm}^2
\ ,
\eeq
where $c_s = \sqrt{\Gamma P/(h\rho_0)}$ is the sound speed, $v_A$ is the Alfv\'en speed, and the
subscript ``cm'' refers to comoving frame values.  To solve this
equation for $\omega/k_1$ in the grid frame, we use $\omega_{\rm cm} = -k_{\mu}u^{\mu}$,
$v_A^2 = b^2/{\cal E}$, and $k_{\rm cm}^2 = K_{\mu}K^{\mu}$, where
${\cal E} = \rho_0h + b^2$ and $K_{\mu} = (g_{\mu\nu} + u_{\mu}u_{\nu})k^{\nu}$
is the part of the wave vector normal to $u^{\mu}$.  Then, for $k^{\mu}$,
one substitutes Eq.~(\ref{kform}) or its $y$-axis or $z$-axis equivalent.

\subsection{Recovery of Primitive Variables}
Having computed ${\bf U}$ at the new timestep, we must use these values to
recover ${\bf P}$, the primitive variables on the new time level. 
This is not trivial because, although the relations ${\bf U}({\bf P})$
are analytic, the inverse relations ${\bf P}({\bf U})$ are not. In general, 
one can do the inversion by numerically solving a system of nonlinear 
algebric equations~\cite{pvarMHD}. 
Here, we discuss the case where the EOS is given by Eq.~(\ref{ideal_P}). 
We need to solve the following
four equations for the variables $\epsilon$ and $u_i$ [c.f. Eqs.~(\ref{momdef}),~(\ref{taudef})]:
\beqn
0 & = & \rho_{\star} h u_i  + \alpha \sg u^0 b^2 u_i - \alpha \sg b^0 b_i - \tilde{S}_i \\ 
0 & = & (\alpha u^0 - 1 + \Gamma \epsilon\alpha u^0)\rho_{\star} + \sg b^2 (\alpha u^0)^2  \nonumber \\
  &   & {} - \sg\left(P+\frac{b^2}{2}\right) - \sg (\alpha b^0)^2 - \tilde\tau \ . 
\label{tau_solve}
\eeqn
In this system, the variables $h$, $u^0$, $b^2$, $b^0$, $b_i$, and $P$ are treated
as functions of the unknown variables
$\epsilon$ and $u_i$, the known set of conserved variables ${\bf U}$, and 
the known metric quantities.
The primitive variables ${\bf P} = (\rho_0, P, v^i, B^i)$ are 
then constructed using the $\epsilon$ and $u_i$ which solve the above system.  The
updated values of $B^i$ are already known from the induction step.  To obtain
the remaining primitive variables, the following set of steps may be used:
\beqn
u^0    & = & \frac{1}{\alpha}(1 + \gamma^{ij}u_i u_j)^{1/2} \\
\rho_0 & = & \frac{\rho_{\star}}{\alpha \sg u^0} \\
P      & = & (\Gamma - 1)\rho_0 \epsilon \\
v^i    & = & \frac{1}{u^0}\gamma^{ij}u_j - \beta^i \ .
\eeqn

As the primitive variable inversion is much simpler without magnetic fields, 
we use a different scheme in such cases.  First, the condition 
$u_{\mu} u^{\mu} = -1$ is rewritten as 
\beq
w^2 = \rho_{\star}^2 + \gamma^{ij}\frac{\tilde{S}_i\tilde{S}_j}{h^2} \ ,
\label{unorm}
\eeq
where $w \equiv \alpha u^0 \rho_{\star}$.  Using the definition
of $\tilde{\tau}$ in Eq.~(\ref{taudef}), one may write $h$ in terms
of $w$ and the conserved variables:
\beq
h = \frac{\Gamma w(\tilde{\tau}+\rho_{\star}) - (\Gamma-1)\rho_{\star}^2}
    {\Gamma w^2-(\Gamma-1)\rho_{\star}^2} \  .
\label{heqn}
\eeq
Substituting this expression into Eq.~(\ref{unorm}) leads to a quartic
equation for $(w-\rho_{\star})$~\cite{footnote:pvar}. We solve this equation
using a standard
polynomial root finder, and then find $h$ by substituting $w$ back into
Eq.~(\ref{heqn}).  The primitive variables can then be constructed according
to the following set of steps:
\beqn
u^0 & = & \frac{w}{\alpha \rho_{\star}} \\ 
\rho_0 & = & \frac{\rho_{\star}^2}{\sg w} \\ 
P & = & \frac{\Gamma-1}{\Gamma}\rho_0 (h-1) \\
v^i & = & \frac{1}{u^0}\gamma^{ij}
          \frac{\tilde{S}_j}{\rho_{\star}h} - \beta^i \ .
\eeqn

\subsection{Low-Density Regions}
If a pure vacuum were to exist anywhere in our computational domain, the MHD
approximation would not apply in this region, and we would have to solve
the vacuum Maxwell equations there (see~\cite{bs03b} for an example).  In many
astrophysical scenarios, however, a sufficiently dense, ionized plasma will exist outside
the stars or disks, whereby MHD will remain valid in its force-free limit.  For the
code tests involving magnetic fields which we will be presenting in this 
paper, there is no such low density region and no special treatment is required.
We do, however, present tests below with unmagnetized, rotating stars.  
For these tests, we do not impose floors on the hydrodynamic variables.  This
is the ``no-atmosphere'' approach used in \cite{dmsb03}.  However, in the 
low-density regions near the surface of the star, we sometimes encounter 
problems when recovering the primitive variables; in particular, the equations 
${\bf U} = {\bf U}({\bf P})$
have no physical solution.  Usually, unphysical ${\bf U}$ are those
corresponding to negative pressure.  At these points, we apply a fix, first
suggested by Font {\it et al}~\cite{fmst00}.  In the system of equations
to be solved, we replace the energy equation~(\ref{tau_solve})
with the adiabatic relation $P = \kappa\rho_0^{\Gamma}$, where $\kappa$ is set
equal to its initial value.  This substitution gaurantees  a
positive pressure.  When magnetic fields are present, the no-atmosphere
approach is not suitable, and a very small positive density must
be maintained outside the stars.  Special techniques for dealing with the
low-density region in MHD calculations have been explored in~\cite{HARM,dVh03}.

\subsection{Constrained Transport}
Unphysical behavior may be expected if the divergence of the magnetic field is
not forced to remain zero.  Thus, {\it constrained} transport schemes have been
designed to evolve the induction equation while maintaining 
$\partial_i \tilde B^i = 0$ to roundoff precision \cite{eh88}.  We use the
flux-interpolated constrained transport (flux-CT) scheme introduced by
T\'oth~\cite{t00}
and used by Gammie {\it et al}~\cite{HARM}.  This scheme involves
replacing the induction equation flux computed at each point with
linear combinations of the fluxes computed at that point and neighboring
points.  The combination assures both that second-order accuracy is
maintained, and $\partial_i\tilde B^i = 0$ is strictly enforced.

\subsection{Black Hole Excision}
Black hole spacetimes are evolved using singularity excision.  This
technique involves removing from the grid a region (the ``excision zone'')
containing the spacetime singularity.  Rather than evolving inside this
region, boundary conditions are placed on the fields immediately outside
the excision zone.  If the region excised is inside the event horizon,
the causal properties of the spacetime will prevent the effects of
excision from contaminating the evolution outside the black hole.  Our
excision zones are spherical and are placed well inside the apparent horizons,
hence well inside the event
horizons.  For details on the excision boundary conditions placed on the
metric fields, see~\cite{ybs02}.  In~\cite{dsy04}, we set the hydrodynamic
variables
equal to zero at the excision boundary (i.e. matter is destroyed when it
hits the excision zone).  With our new code, we find that this excision
boundary condition for the fluid variables is still adequate in the
absence of magnetic fields.  When magnetic fields are present, however,
it can become problematic.  Therefore, we now set the MHD variables on
the excision boundary by linearly extrapolating the primitive variables
along the normal to the excision surface.

\section{Code Tests}

\label{codetests}

\subsection{Unmagnetized Relativistic Stars}

\newcommand{\stara}{A}
\newcommand{\starb}{B}
\newcommand{\starc}{C}

\begin{table}
\caption{Rotating Equilibrium Stars ($n = 1$, $\kappa = 1$)$^a$.}
\begin{tabular}{cccccccc}
\hline \hline  
 Star & $M\ {}^b$ &
 $\left.\right.\ R_{\rm eq}{}^c\ \left.\right.$ & $R_c{}^d$
 & ${\cal R}\ {}^e$ & $J/M^2$ & $T/|W|{}^f$ & $\Omega_p/\Omega_{\rm eq}{}^g$
 \\ \hline  
\stara\ & 0.170 & 0.540 & 0.881 & 0.88 & 0.35 & 0.032 & 1.00 \\
\starb\ & 0.171 & 0.697 & 0.780 & 0.87 & 0.34 & 0.031 & 1.00 \\
\starc\ & 0.279 & 1.251 & 1.613 & 0.30 & 1.02 & 0.230 & 2.44 \\
\hline \hline  
\label{table:stars}
\end{tabular}

\raggedright

${}^a$ The maximum ADM mass for a nonrotating $n = 1$, $\kappa = 1$ polytrope
       is $M_{\rm max} = 0.164$.

${}^b$ ADM mass 

${}^c$ coordinate equatorial radius

${}^d$ circumferential radius at the equator

${}^e$ ratio of polar to equatorial coordinate radius

${}^f$ ratio of rotational kinetic to gravitational potential energy

${}^g$ ratio of polar (central) to equatorial angular velocity

\end{table}

\begin{figure}
\includegraphics[width=8cm]{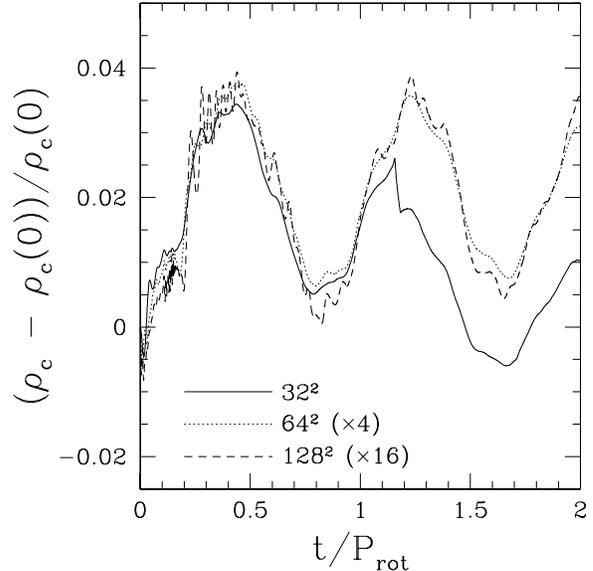}
\caption{Relative error in the central rest-mass density 
for uniformly rotating star~\stara. 
The error is plotted for three axisymmetric runs, adopting equatorial 
symmetry (with resolutions 
$32^2$, $64^2$, and $128^2$), and the curves are scaled for second-order
convergence. All runs are performed with the standard 
hydrodynamic scheme (HLL with PPM$^+$).  Outer boundaries are placed
at $7.1M$.}
\label{fig:conv}
\end{figure}

In this section, we test the ability of our GRMHD code to handle 
rotating relativistic stars without magnetic fields.
For initial data, we take a perfect fluid with a polytropic equation of state
$P = \kappa\rho_0^{1+1/n}$, with $n = 1$, and we choose our units such
that $\kappa = 1$. (For a description of the code used to generate these rotating
equilibrium stars, see~\cite{cst92}.  Eqs.~(15)--(23) of~\cite{cst92} give the scaling
relations to arbitrary $\kappa$.) 
We evolve the three rotating polytropes described in
Table~\ref{table:stars}.  We adopt equatorial symmetry in all cases. 
We note that stars \stara, \starb, and
\starc\ in Table~\ref{table:stars} correspond, respectively, to stars
C, D, and E in Table I of~\cite{dmsb03}. 
First, we demonstrate convergence for axisymmetric evolutions of star \stara,
a uniformly
rotating, stable star.  This star is known to be secularly stable by the turning-point
theorem~\cite{fis88}, and it is known to be dynamically stable from previous
numerical simulations~\cite{dmsb03}.  Thus, we should find that the system
maintains equilibrium when evolved in our code. 
In Fig.~\ref{fig:conv}, we show the error in the central density for three 
short runs with star~\stara\ at different resolutions.  This demonstrates that our 
standard method for hydrodynamics (HLL fluxes with PPM$^+$ reconstruction) 
leads to second-order convergence.  

\begin{figure}
\includegraphics[width=8cm]{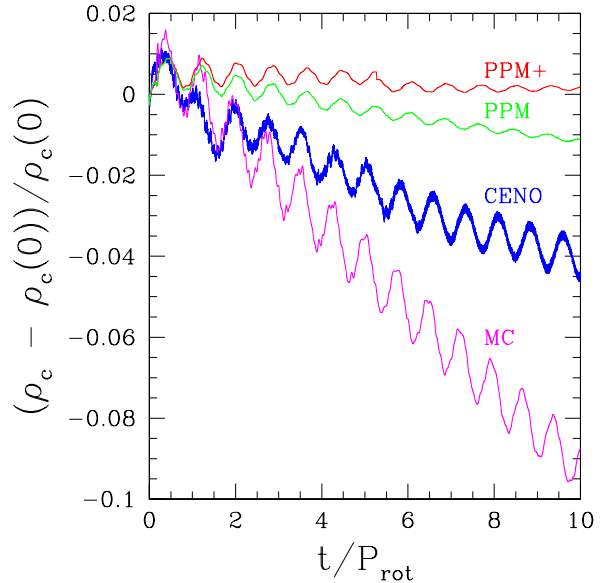}
\caption{Normalized error in the central rest-mass density for 
star \stara\ with different reconstruction methods. 
All runs are axisymmetric and equatorially symmetric 
with a resolution of $64^2$ and outer boundaries at $7.1M$. 
With PPM, the drift in the
central density is strongly reduced with respect to the MC and CENO results. 
With PPM$^+$, the central density drift disappears
almost entirely.}
\label{fig:recon}
\end{figure}

Evolutions of star~\stara\ using several different reconstruction methods are 
compared in Fig.~\ref{fig:recon}.  Reconstruction
with the MC limiter leads to a downward drift in the
central rest-mass density ($\sim 10 \%$ in $10 P_{\rm rot}$, where $P_{\rm rot}$ is
the rotation period).  A similar
drift has been seen in simulations using other codes~\cite{fgimrssst02,fsk00}. 
We find that the drift converges to zero faster than second order
as the resolution is increased.  CENO reconstruction gives a slight improvement,
but introduces high frequency oscillations in the central density.  In
Fig.~\ref{fig:recon}, these 
oscillations are not individually distinguishable, 
but instead make the CENO line appear thicker than the others.  
These oscillations are an artifact of the coordinate
singularity near the axis in cylindrical coordinates. (We do not see
the oscillations in 3D runs with CENO.) 
The high frequency oscillations can be removed by adding high-order 
dissipation. We note, however, that the other reconstruction methods 
represented in Fig.~\ref{fig:recon} do not display high frequency oscillations
and thus do not require such fixes.  The best results are achieved with PPM 
and PPM$^+$ reconstruction.  Much of the drift in the central density vanishes
when standard PPM is used, which is consistent with the result reported 
in~\cite{fsk00}.  However, with PPM$^+$, the drift is eliminated almost entirely.

\begin{figure}
\vskip 1cm
\includegraphics[width=8cm]{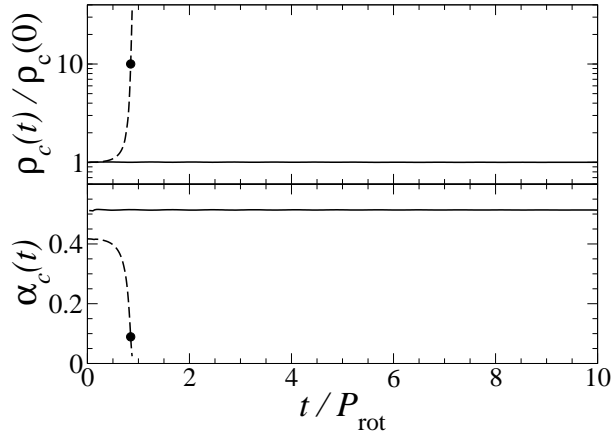}
\vskip 0.5cm
\caption{Axisymmetric evolution of uniformly rotating stars. Star~\stara\
(solid lines) is stable, while star~\starb\ (dashed lines) is unstable to
collapse. The upper window shows the central density normalized to its initial
value, while the lower gives the central lapse.  The solid dot indicates the
first appearance of an apparent horizon during the collapse of star~\starb.}
\label{fig:starstability}
\end{figure}

Next we check the ability of our code to distinguish radially stable
from radially unstable stars.  We consider two uniformly rotating
stars, stars \stara\ and \starb, which are members of a constant angular momentum
sequence, $J=0.01$ in our $G = c = \kappa = 1$ units.  The $J = 0.01$ sequence has a
turning-point at central rest-mass density $\rho_c^{\rm crit} = 0.31$, which has the maximum
mass $M_{\rm max} = 0.172$ for the sequence.  For a sequence of
uniformly rotating stars, this turning point marks the onset of
secular, not dynamical, radial instability~\cite{fis88}, but prior numerical
simulations~\cite{dmsb03} have found the point of onset of dynamical
instability to be very close to the point of onset of secular instability.
We pick two similar stars on either side of the onset of secular
instability: star \stara\ with initial central rest-mass density $\rho_c(0) = 0.24$
on the stable branch and Star \starb\ with $\rho_c(0) = 0.37$ on the unstable
branch.  In Fig.~\ref{fig:starstability}, we see that the code correctly
finds star \stara\ to be stable and star \starb\ to be unstable.

\vskip 1cm
\begin{figure}
\includegraphics[width=8cm]{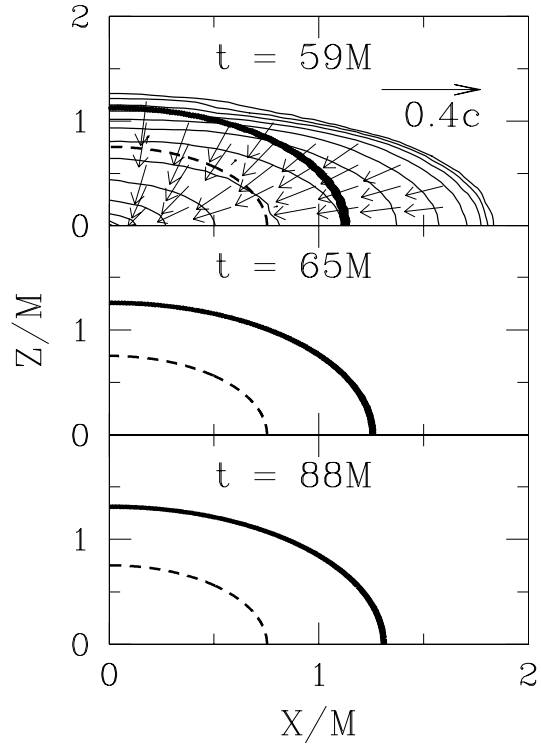}
\vskip 0.5cm
\caption{
  Snapshots of the rest density contours and the
  velocity field $(v^x,v^z)$ in the meridional plane during the axisymmetric
  collapse of uniformly rotating star \starb\ to a Kerr black hole. The contour lines are
  drawn for $\rho_0=10^{-(0.3j+0.09)} \rho_{0}^{Max} $ for $j=0,1,..,12$, 
  where $\rho_{0}^{Max}$ is the maximum of $\rho_0$ at the time of excision. 
  The thick dashed curve marks the excision zone.  The thick solid curve is the
  apparent horizon.  We show the system at the time of excison (upper panel),
  at the time at which the last of the matter falls into the excision region
  (middle panel), and at a late time just prior to the termination of the integration
  (lower panel). }
\label{fig:excision}
\end{figure}

Star \starb\ collapses to a black hole.  Without excision, the extreme density and spacetime
curvature at the center of the collapsing star cause the code to crash
shortly after the formation of an apparent horizon which envelops some,
but not all, of the star. 
The evolution can be continued by excising a region inside the horizon. 
When we do this, we find that all of the matter falls into the hole within
a few $M$ of the time excision is introduced, leaving a vacuum Kerr
black hole with roughly the same $M$ and $J$ as the initial star \starb. 
We then continue to evolve for another $30M$. 
We find that the hole's angular momentum, computed as the sum of surface and
volume integrals, decreases
with time, and this angular momentum drift limits the length of time that our
evolution remains accurate.  Comparable angular momentum loss was also present
in~\cite{dsy04}.  Since this drift
appears after most of the matter has fallen into the excision region, the
source of the error resides in the evolution of the BSSN variables.  Since
we use the same algorithm to evolve the metric as in~\cite{dsy04}, it is
not surprising that the $J$-drift has the same magnitude.  We simulated the collapse
of star \starb\ on both $64^2$ and $128^2$ grids, and we found, as in~\cite{dsy04},
that the $J$-drift converges to zero with increasing resolution.  In Fig.~\ref{fig:excision},
we show snapshots from our post-excision evolution on the $128^2$ grid~\cite{footnote:ex}. 
To check that the final geometry corresponds to a stationary Kerr spacetime, we confirm
that the area and the equatorial and polar circumferences of the apparent horizon agree
with the appropriate values for a Kerr black hole with the $M$ and $J$
of our spacetime to better than 1\% (see~\cite{dsy04} for details).

\begin{figure}
\includegraphics[width=8cm]{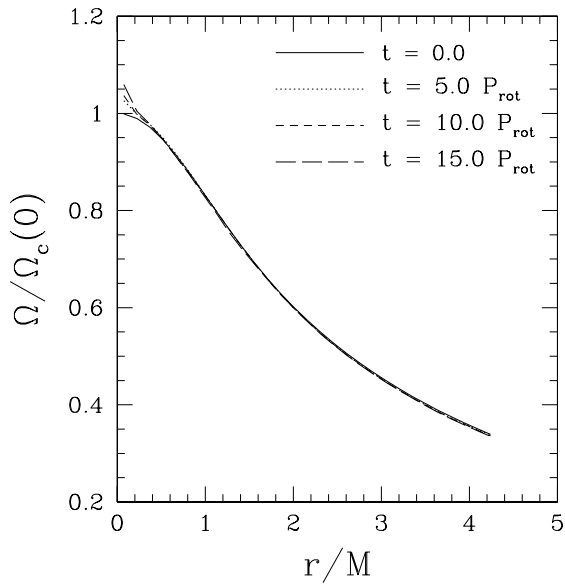}
\caption{Snapshots of the angular velocity profile for an evolution
of the differentially rotating star~\starc\ in axisymmetry ($48^2$) 
with outer boundaries at $7.1 M$.  
The profile is well-maintained in
the bulk of the star for over 15 central rotation periods 
($P_{\rm rot}$).}
\label{fig:omprof}
\end{figure}

We now demonstrate the ability of our code to handle differential
rotation in both 2D and 3D by considering the evolution of star \starc,
a stable, hypermassive, differentially rotating star. 
We evolved this star in axisymmetry with PPM$^+$ using
a fairly low resolution of $48^2$ zones and outer boundaries at $7.1 M$.  
Figure~\ref{fig:omprof} shows the angular velocity profile at 
several times during this evolution, and demonstrates that our
code correctly maintains the differential rotation.  Slight errors in the 
angular velocity arise near the origin.  The central angular velocity
is particularly susceptible to error, as its calculation involves dividing
the local azimuthal velocity by the (small) radius.  During the first 
$15 P_{\rm rot}$ of this evolution (where $P_{\rm rot}$ refers to the 
central rotation period), the central density remains within
$5 \%$ of its initial value and the ADM mass is conserved to within
$0.5 \%$, even at this low resolution.  (The angular momentum computed
as a volume integral over all space is
conserved exactly in our code in axisymmetry.)  
During this period, the normalized Hamiltonian and momentum 
constraints (see~\cite{dmsb03} for definition) are $\sim 1~\%$.

\begin{figure}
\includegraphics[width=8cm]{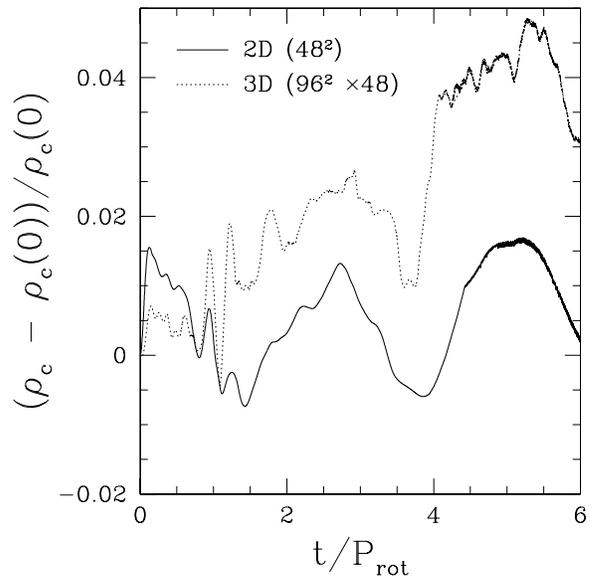}
\caption{Fractional error in the central density versus time
 for evolutions of star~\starc\ in 2D and 3D.  The 3D run was performed in 
equatorial symmetry with resolution $96^2 \times 48$ with outer 
boundaries at $7.1M$.  The axisymmetric run has the same spatial
resolution (grid cell size) as the 3D case, and thus employs a $48^2$
grid with boundaries at $7.1M$.  The errors in the central density are 
of the same order for both runs.}
\label{fig:2d3d}
\end{figure}

Next, we compare the axisymmetric run of star~\starc\ described above with
an equivalent 3D run in equatorial symmetry.  The 3D run was performed
with a $96^2 \times 48$ grid in $(x,y,z)$ and outer boundaries at $7.1M$, 
so that the grid cell size is the same as in the 2D run.  The deviations 
in the central density for these two runs are compared in Fig.~\ref{fig:2d3d},
which shows that the axisymmetric and full 3D runs have comparable errors.
For $t \lesssim 6 P_{\rm rot}$, the two runs are similar and the 
angular velocity profiles agree very well.  After this time, however, 
the star in the 3D run begins to move away from the center of the grid, 
eventually making contact with the outer grid boundary.  This is due
to accumulated error in the linear momentum and is a well known problem
associated with evolutions of stars in equatorial symmetry~\cite{nct00}. 
In our simulations, the effect can be reduced by improving spatial
resolution.  By comparing runs of star~\starc\ on $64^2 \times 32$,
$96^2 \times 48$, and $168^2 \times 84$ grids, we find that the movement
of the center of mass converges to zero at third-order in spatial resolution. 
We note that the drift can be removed at any resolution by employing
$\pi$-symmetry, so that the symmetry boundary conditions tie
the star to the center of the grid.    

\subsection{Minkowski Spacetime MHD Tests}
\label{kom_tests}

\begin{figure}
\includegraphics[width=8.5cm,height=10cm]{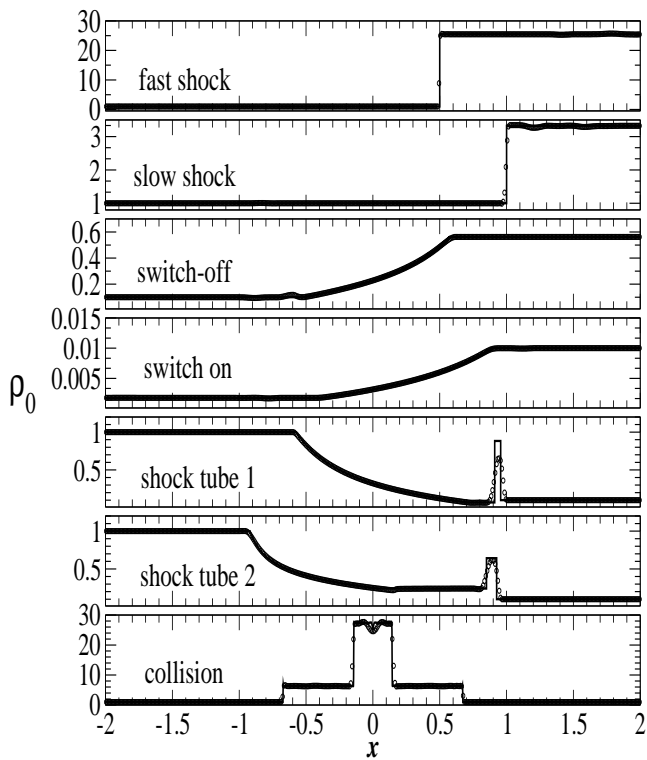}
\caption{Density profiles for the nonlinear wave tests
at $t=t_{\rm final}$ (see Table~\ref{tab:1Dtests}). Symbols denote
data from numerical simulations with resolution $\Delta x =0.01$.
Solid lines in the upper 6 panels denote the exact 
solutions~\cite{ko_exact}. The solid line in the last panel denotes a
numerical simulation with higher resolution, $\Delta x = 10^{-3}$.}
\label{fig:ko_rho}
\end{figure}

\begin{figure}
\includegraphics[width=8.5cm,height=10cm]{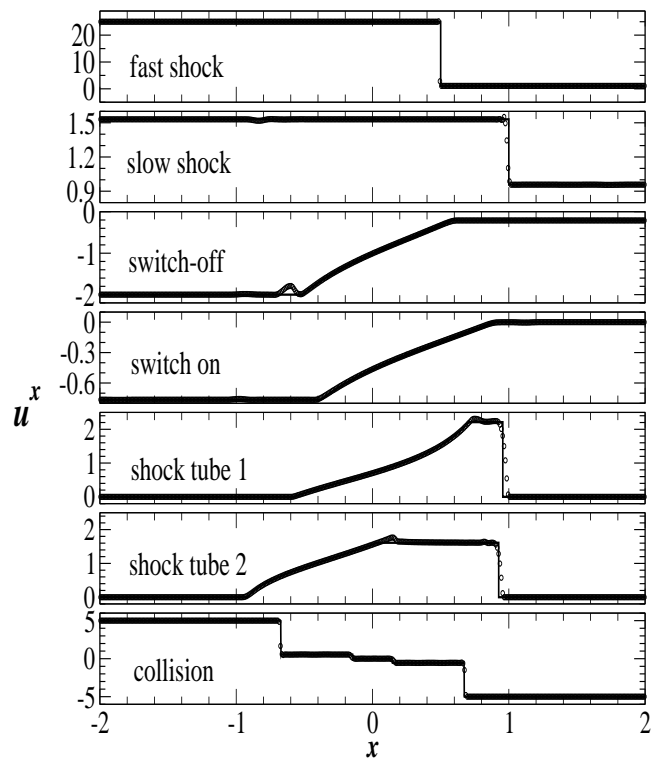}
\caption{Velocity profiles ($u^x$) for the nonlinear wave tests
at $t=t_{\rm final}$. Symbols denote
data from numerical simulations with resolution $\Delta x = 0.01$.
Solid lines in the upper 6 panels denote
the exact solutions~\cite{ko_exact}. The solid line in the last panel denotes
a numerical simulation with higher resolution, $\Delta x =10^{-3}$.}
\label{fig:ko_ux}
\end{figure}

\begin{figure}
\vskip 0.5cm
\includegraphics[width=8cm]{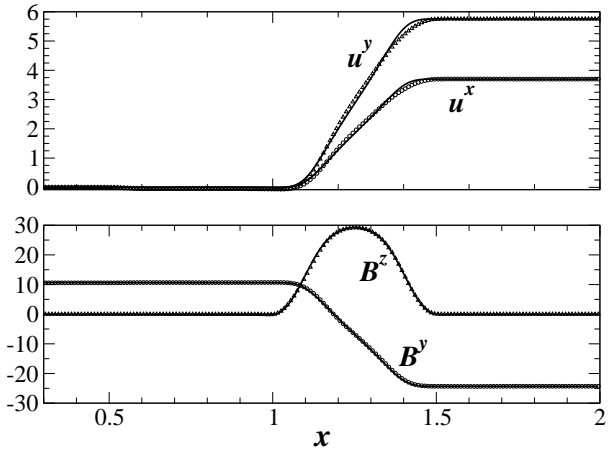}
\caption{Nonlinear Alfv\'en wave test. Symbols are simulation results
with resolution $\Delta x=0.01$ and solid lines are the exact solution. 
The profiles are shown at time $t=t_{\rm final}=2.0$. Our computational 
domain is $x \in (-2,2)$. We only show the region $0.3 \leq x \leq 2.0$ 
in this graph.}
\label{fig:alfven}
\end{figure}

Komissarov~\cite{k99} has proposed a suite of challenging 
one-dimensional tests of nonlinear, relativistic MHD waves in Minkowski 
spacetime. Most of the tests 
(except the nonlinear Alfv\'en wave test) start with discontinuous 
initial data at $x=0$ (see Table~\ref{tab:1Dtests}), with homogeneous
profiles on either side. We integrate
the MHD equations from $t=0$ to $t=t_{\rm final}$, where $t_{\rm final}$ 
is specified in
Table~\ref{tab:1Dtests} for each case. The gas satisfies a $\Gamma$-law EOS 
with $\Gamma=4/3$. In all the cases, 
our computational domain is $x \in (-2,2)$. Our standard resolution 
is $\Delta x = 0.01$ (400 grid points). We are 
able to integrate all the cases using the MC resconstruction 
scheme and with a Courant factor of 0.5. Thus, the number of timesteps
for a given test is $t_{\rm final}/(0.5\Delta x) = 200 t_{\rm final}$.  
With PPM resconstruction scheme,
we need to lower the Courant factor to 0.4 for the fast shock test.
We obtain slightly better results with PPM resconstruction scheme
for the shock tube tests 1 and 2. The two resconstruction schemes 
give comparable results for the other tests. Here we present the 
simulations using the MC resconstruction scheme. 
Figures~\ref{fig:ko_rho}--\ref{fig:alfven}
compare our simulation results (symbols) with the expected
results (solid lines)~\cite{ko_exact}.
Our numerical results are similar to those reported recently for other 
codes~\cite{k99,HARM}. Below, we briefly discuss each of the cases we studied.

\begin{table*}
\caption{Initial states for one-dimensional MHD tests.${}^{\rm a}$}
\begin{tabular}{c  c  c  c}
\hline \hline
 Test & Left state & Right State & $t_{\rm final}$ \\ 
\hline \hline
  Fast Shock & $u^i=(25.0,0.0,0.0)$ & $u^i=(1.091,0.3923,0.00)$ & 2.5 \\
 ($\mu=0.2{}^{\rm b}$) & $B^i/\sqrt{4\pi}=(20.0,25.02,0.0)$ & 
 $B^i/\sqrt{4\pi}=(20.0,49.0,0.0)$ &  \\
  & $P=1.0$, $\rho_0=1.0$ & $P=367.5$, $\rho_0=25.48$ & \\
\hline
  Slow Shock & $u^i=(1.53,0.0,0.0)$ & $u^i=(0.9571,-0.6822,0.00)$ & 2.0 \\
 ($\mu=0.5{}^{\rm b}$) & $B^i/\sqrt{4\pi}=(10.0,18.28,0.0)$ & 
 $B^i/\sqrt{4\pi}=(10.0,14.49,0.0)$ & \\
  & $P=10.0$, $\rho_0=1.0$ & $P=55.36$, $\rho_0=3.323$ & \\
\hline
  Switch-off Fast & $u^i=(-2.0,0.0,0.0)$ & $u^i=(-0.212,-0.590,0.0)$ & 1.0 \\
  Rarefaction & $B^i/\sqrt{4\pi}=(2.0,0.0,0.0)$ & 
  $B^i/\sqrt{4\pi}=(2.0,4.71,0.0)$ & \\
  & $P=1.0$, $\rho_0=0.1$ & $P=10.0$, $\rho_0=0.562$ & \\
\hline
  Switch-on Slow & $u^i=(-0.765,-1.386,0.0)$ & $u^i=(0.0,0.0,0.0)$ & 2.0 \\
  Rarefaction & $B^i/\sqrt{4\pi}=(1.0,1.022,0.0)$ & 
 $B^i/\sqrt{4\pi}=(1.0,0.0,0.0)$ & \\
 & $P=0.1$, $\rho_0=1.78\times 10^{-3}$ & $P=1.0$, $\rho_0=0.01$ & \\
\hline
  Shock Tube 1 & $u^i=(0.0,0.0,0.0)$ & $u^i=(0.0,0.0,0.0)$ & 1.0 \\
  & $B^i/\sqrt{4\pi}=(1.0,0.0,0.0)$ & $B^i/\sqrt{4\pi}=(1.0,0.0,0.0)$ & \\
  & $P=1000.0$, $\rho_0=1.0$ & $P=1.0$, $\rho_0=0.1$ & \\
\hline
  Shock Tube 2 & $u^i=(0.0,0.0,0.0)$ & $u^i=(0.0,0.0,0.0)$ & 1.0 \\
 & $B^i/\sqrt{4\pi}=(0.0,20.0,0.0)$ & $B^i/\sqrt{4\pi}=(0.0,0.0,0.0)$ & \\
 & $P=30.0$, $\rho_0=1.0$ & $P=1.0$, $\rho_0=0.1$ & \\
\hline
  Collision & $u^i=(5.0,0.0,0.0)$ & $u^i=(-5.0,0.0,0.0)$ & 1.22 \\
  & $B^i/\sqrt{4\pi}=(10.0,10.0,0.0)$ & $B^i/\sqrt{4\pi}=(10.0,-10.0,0.0)$ & \\
  & $P=1.0$, $\rho_0=1.0$ & $P=1.0$, $\rho_0=1.0$ & \\
\hline
Nonlinear Alfv\'en wave${}^{\rm c}$ & $u^i=(0.0,0.0,0.0)$  & 
$u^i=(3.70,5.76,0.00)$ & 2.0 \\
($\mu=0.626{}^{\rm b}$) & $B^i/\sqrt{4\pi}=(3.0,3.0,0.0)$ & 
$B^i/\sqrt{4\pi}=(3.0,-6.857,0.0)$ & \\
 & $P=1.0$, $\rho_0=1.0$ & $P=1.0$, $\rho_0=1.0$ & \\
\hline \hline
\end{tabular}
\vskip 12pt
\begin{minipage}{12cm}
\raggedright
${}^{\rm a}$ {In all cases, the gas satisfies the $\Gamma$-law EOS with 
$\Gamma = 4/3$. For the first 7 tests, the left state refers to $x<0$ 
and the right state, $x>0$.} \\
${}^{\rm b}$ {$\mu$ is the speed at which the wave travels} \\
${}^{\rm c}$ {For the nonlinear Alfv\'en wave, the left and right 
states are joined by a continuous function. See~\cite{k97} 
or Appendix~\ref{app:alfven} for details.} \\
\end{minipage}
\label{tab:1Dtests}
\end{table*}

{\it Fast and slow shocks} In these tests, the initial MHD variables on 
the left ($x<0$) and right ($x>0$) satisfy the special relativistic 
Rankine-Hugoniot jump conditions for MHD shocks~\cite{ma87}. 
As a result, the discontinuity simply travels with a certain speed 
$\mu$ without changing its pattern. The fast shock is the most 
relativistic case of all the tests. In the shock frame, the 
Lorentz factor of the upstream flow is $u^0 \approx 25$. The shock 
moves with a speed $\mu=0.2$. The slow shock is not as strong and 
it moves with a faster speed ($\mu=0.5$). Our simulations for these
two tests agree quite well with the exact solutions. In the slow 
shock test, we see small oscillations (due to numerical artifacts) 
in density on the right side of the shock (see Fig.~\ref{fig:ko_rho}). 
This numerical artifact is also seen in simulations with other
codes~\cite{k99,HARM}. We have performed simulations on 
these two cases using different resolutions and found that the 
errors converge to first order in $\Delta x$, which is 
expected for problems with discontinuities in the computational 
domain.

{\it Switch-on/off rarefaction} In these tests, the left and right 
states are connected by a rarefaction wave at $t>0$. The tests 
become more challenging when the tangential component of the 
magnetic field (i.e., $B^y$) is switched on/off when going 
from the right state to the left state. The exact solutions 
are obtained by integrating a system of ordinary differential 
equations (see e.g.,~\cite{k99,c70}). Our simulation results agree with 
the exact solutions very well, except that we see numerical 
artifacts near the trailing edge of the rarefaction wave 
in the switch-off test. We also observe a 
small oscillation (not visible on the scale shown in 
Figs.~\ref{fig:ko_rho} and~\ref{fig:ko_ux}) near the leading 
edge of the rarefaction wave in the switch-on test. These 
numerical artifacts are also seen in simulations with other
codes~\cite{k99,HARM}. As explained in~\cite{k99}, the oscillation
results from perturbations created by numerical dissipation during 
the initial stage when the wavefront is very steep. The perturbations 
propagate across the main wave and then separate from it.

{\it Shock tubes 1 and 2} The initial left and right states are 
given in Table~\ref{tab:1Dtests}. At time $t>0$, the left and right 
states are connected by a rarefaction wave, a contact discontinuity 
and a shock wave. The exact solution can be computed using a method 
similar to~\cite{t86}. In the shock tube 1 test, the solution 
consists of a thin layer of shocked gas, which is poorly resolved 
in our simulation and has a wrong value of shell density. The thin 
layer is covered by only 5 grid points with our resolution. We found that 
the correct density is obtained in higher resolution simulations
in which $\Delta x \lesssim 0.0035$, which provides $\gtrsim$ 12 grid points across 
the thin layer. Our results in these two tests are comparable to those 
reported in~\cite{k99,HARM}.

{\it Collision} In this test, the flows on both sides travel with 
equal speed but in opposite directions. The tangential component of the 
magnetic field is also equal in magnitude but opposite in direction. 
We do not have the exact solution for this test. 
Thus, we compare our lower-resolution ($\Delta x =0.01$) simulation 
with a high-resolution one ($\Delta x = 10^{-3}$). 
The lower-resolution 
simulation results are qualitatively the same as the results reported 
in~\cite{HARM}, but not as good as the results of Komissarov~\cite{k99}, 
who uses a more sophisticated Riemann solver. 

\begin{figure}
\vskip 0.5cm
\includegraphics[width=8cm]{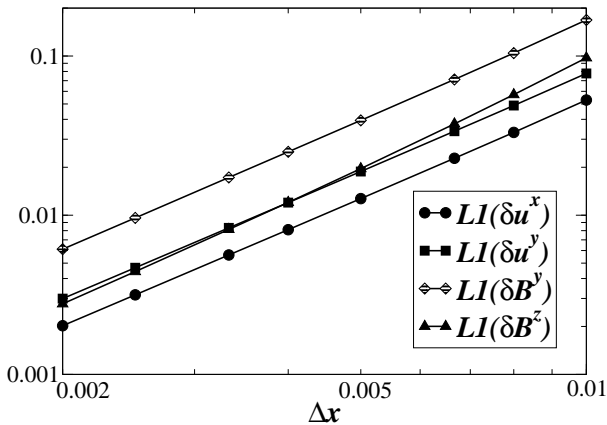}
\caption{L1 norms of the errors in $u^x$, $u^y$, $B^y$ and $B^z$ for 
the nonlinear Alfv\'en wave test at $t=t_{\rm final}=2.0$.
This log-log plot shows that the 
L1 norms of the errors in $u^x$, $u^y$ and $B^y$ are proportional 
to $(\Delta x)^2$, and are thus second-order convergent. The error in 
$B^z$ goes as a slightly higher power of $\Delta x$.}
\label{fig:l1_alfven}
\end{figure}

{\it Nonlinear Alfv\'en Wave} The initial data for this test 
are qualitatively different from the other seven tests. The 
left ($x<-W/2$) and right ($x>W/2)$ states are separated by 
a width $W=0.5$ at $t=0$. The two states are joined by 
continuous functions in the region $x \in (-W/2,W/2)$
at $t=0$. The details of the setup of initial data can be found 
in~\cite{k97}, which we summarize in Appendix~\ref{app:alfven}. 
The pattern should simply move with a constant speed $\mu=0.626$. 
Figure~\ref{fig:alfven} shows the simulation results (symbols) 
and exact solution (solid lines). The simulation results are 
again similar to~\cite{k99}. Since there are no discontinuities 
in this problem, we expect the errors to converge at second order
in $\Delta x$. To demonstrate this, we consider 
a grid function $g$ with error $\delta g = g - g^{\rm exact}$. 
We calculate the L1 norm of $\delta g$ (the ``average'' of 
$\delta g$) by summing over every grid point $i$:
\beq
  L1(\delta g) \equiv \Delta x \sum_{i=1}^N |g_i-g^{\rm exact} (x_i)| \ ,
\eeq
where $N \propto 1/\Delta x$ is the number of grid points. 
Figure~\ref{fig:l1_alfven} shows the L1 norms of the errors 
in $u^x$, $u^y$, $B^y$ and $B^z$ at $t=t_{\rm final}=2.0$. 
We find that the errors in $u^x$, $u^y$ and $B^y$ converge 
at second order in $\Delta x$. The error in $B^z$ converges 
at slightly better than second order in $\Delta x$.

\subsection{Curved Background Tests: \\ \ \ Relativistic Bondi Flow}
\label{subsection:bondi}

Next, we test the ability of our code to accurately evolve the
relativistic MHD equations in the strong gravitational field near a
black hole.  Specifically, we check its ability to maintain stationary,
adiabatic, spherically symmetric accretion onto a Schwarzschild black hole,
in accord with the relativistic Bondi accretion solution~\cite{st83}. 
It has been shown that the relativistic Bondi solution is unchanged in
the presence of a divergenceless radial magnetic field~\cite{dVh03}. 
The advantages of this test are that it involves strong gravitational 
fields and relativistic
flows, and that there exists an analytic solution with which to test our 
results. We write the metric in Kerr-Schild (ingoing Eddington-Finklestein)
coordinates; in this way, all the variables are
well behaved at the horizon (``horizon penetrating''), and the excision radius can be placed inside
the event horizon.  We begin by holding the metric field
variables fixed in order to prevent the black hole from
growing due to accretion.  With the metric fixed, the flow is exactly stationary in the
continuous limit.  When evolved with a finite-difference code,
discretization errors will cause small deviations in the flow from
its initial state.  These deviations should converge to zero as resolution
is increased.  Eventually, the system may settle down to an equilibrium
solution of the descretized hydrodynamic equations.  To diagnose the
behavior of our code, we introduce two variables.  The deviation of the
fluid configuration from the analytic Bondi solution we measure by
$\delta\rho_{\star}$, the L1 norm in 3D of
$\left|\rho_{\star}-\rho_{\star}^{\rm exact}\right|$ (where 
$\rho_{\star}^{\rm exact}$ is the analytic value of $\rho_{\star}$), normalized 
by the rest mass: 
\beq
\delta \rho_{\star} \equiv 
\frac{\Delta x\Delta y\Delta z \sum_{i,j,k}|\rho_{\star i,j,k}-
\rho_{\star}^{\rm exact}(x_i,y_j,z_k)| }
{\Delta x\Delta y\Delta z \sum_{i,j,k} 
\rho_{\star}^{\rm exact}(x_i,y_j,z_k)} \ .
\eeq
To measure the settling down of the solution to a
numerical equilibrium, we monitor $\Delta\rho_b$, the L2 norm of
$\dot\rho_b\Delta t$:
\beq
\Delta\rho_b \equiv \Delta t \Delta x \Delta y \Delta z \left[ \sum_{i,j,k} 
(\dot\rho_b)_{i,j,k}^2 \right]^{1/2} \ .
\eeq  
The quantities $\delta\rho_{\star}$ and $\Delta\rho_b$
were chosen because they correspond to the diagnosics used to monitor
Bondi accretion test problems in~\cite{HARM} and~\cite{dsy04}, respectively.

\begin{figure}
\includegraphics[width=8cm]{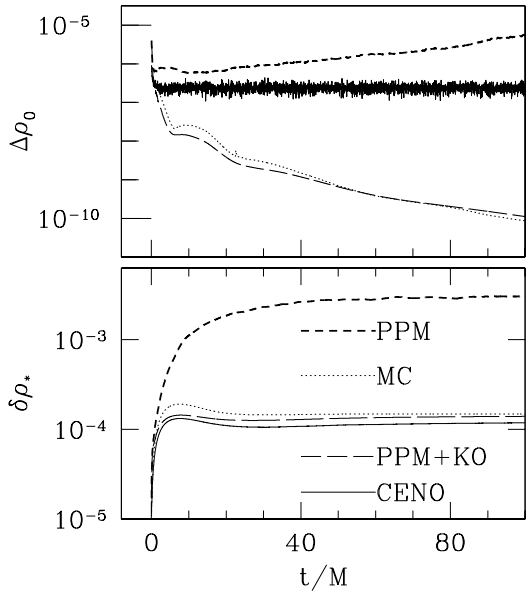}
\caption{Unmagnetized Bondi accretion onto a Schwarzschild black hole. 
Four different methods are compared:  MC reconstruction, CENO reconstruction,
and PPM reconstruction, both with and without Kreiss-Oliger (KO) dissipation. 
Each run was performed on a $64^2$ grid using axisymmetry.}
\label{fig:bondi1}
\end{figure}

For this test, we evolve the same configuration used by~\cite{hsw84},
\cite{dVh03}, and~\cite{HARM}.  The sonic radius is at Schwarzschild (areal)
radius $r_s = 8M$, the accretion rate
is $\dot M = 1$, and the equation of state is $\Gamma = 4/3$.  As
in~\cite{HARM}, we set our excision radius at $r_{\rm ex} = 1.9M$ (the 
horizon is at $2M$) and evolve for $100M$.  (We find that the system 
settles to equilibrium long before
$100M$.)  We place outer boundaries at $10M$, at which point the analytic
values of the conserved variables are imposed. 

First, we evolve this accretion flow in the absence of a magnetic field.  
In Figure~\ref{fig:bondi1}, we show the results for an axisymmetric grid 
of $64^2$ using various numerical
techniques.  We find that using MC reconstruction gives much better results
than using PPM. 
This is probably due to the larger numerical diffusivity in the MC scheme which
stabilizes spurious numerical oscillations. PPM can be ``corrected,'' however, 
by adding a
small dissipation.  In~\cite{hln05}, the oscillations are removed by shifting the
numerical spatial stencil in supersonic flow.  We have instead
addressed the problem by adding a small Kreiss-Oliger dissipation~\cite{ko73}. 
We find that PPM with Kreiss-Oliger dissipation performs as well as MC for
this problem (see Fig.~\ref{fig:bondi1}).
We have also evolved an accretion flow with small $\dot M$ while allowing
the metric to evolve.  We find that the flow is stable, and that, as expected,
the irreducible mass of the black hole slowly increases~\cite{fn:bondi}. 
The MHD variables
remain near the Bondi equilibrium initial values until the black hole grows
appreciably.

\begin{figure}
\includegraphics[width=8cm]{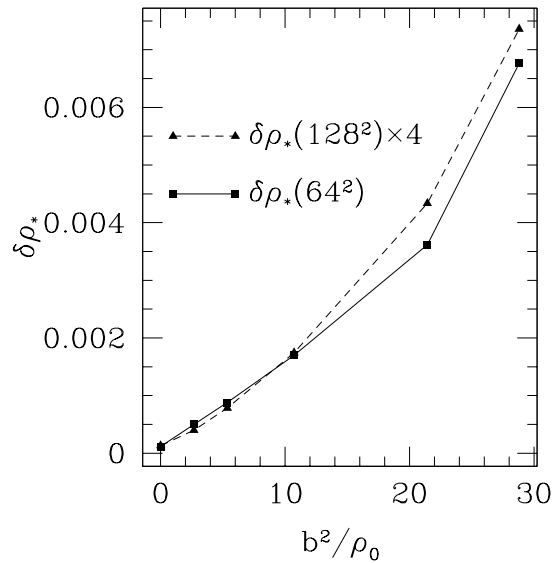}
\caption{$\delta\rho_{\star}$ after $100 M$ for various initial values
of $b^2/\rho_0|_{r=2M}$.  PPM with Kreiss-Oliger dissipation is used in
each run.  Results from $64^2$ and $128^2$ grid runs are compared.  The
$\delta\rho_{\star}$ found with the $128^2$ grid is multiplied by 4 to
allow scaling to be checked for second order convergence.}
\label{fig:bondi2}
\end{figure}

Next, we evolve with a radial magnetic field.  In the continuum limit,
the magnetic forces cancel exactly.  However, the cancellation will not
be exact in a finite-difference code, and this test can be quite difficult
for a GRMHD code when the magnetic field is strong.

In Fig.~\ref{fig:bondi2}, we plot the error, measured by $\delta\rho_{\star}$
after $100M$ of evolution, for 2D (axisymmetric) runs with various
values of $b^2/\rho_0$ at the horizon.  We use the PPM reconstruction
method with Kreiss-Oliger dissipation for each run.  In order to test convergence,
we use both $64^2$ and $128^2$ grids.  We are able to evolve with magnetic fields
$b^2/\rho_0|_{r=2M} \lesssim 30$.  Stronger radial fields quickly crash the code. 
For $b^2/\rho_0|_{r=2M} \lesssim 5$, we find that $\delta\rho_{\star}$ settles quickly
to a final value.  For larger $b^2/\rho_0|_{r=2M}$, $\delta\rho_{\star}$ does not settle
as well, but the results are still second-order convergent after $100M$.
Evolving with MC gives better ``settling'' behavior for
$5\lesssim b^2/\rho_0|_{r=2M} \lesssim 30$,
but it is still not possible to evolve flows with $b^2/\rho_0|_{r=2M} \gtrsim 30$. 
Gammie, McKinney, and T\'oth~\cite{HARM} are able to evolve with much higher
$b^2/\rho_0|_{r=2M}$. 
This is probably due to their use of a spherical-polar coordinate grid, which is
better adapted to the spherical symmetry of this problem than our cylindrical grid.

We have also evolved the Bondi flow in three dimensions on $64^3$ grids using
octant symmetry.  We find that we can maintain equilibrium flow for
$b^2/\rho_0|_{r=2M} \lesssim 10$ in 3D.

\subsection{Dynamical Background Tests: Gravitational Wave-Induced MHD Waves}

To test the capability of our code to handle dynamical gravitational 
and MHD fields simultaneously, we consider a gravitational wave oscillating in an initially 
homogeneous, uniformly magnetized fluid. The gravitational wave will, in 
general, induce Alfv\'en and magnetosonic 
waves~\cite{moortgat03,moortgat04,kallberg04}.  
In a companion paper \cite{dlss05b} (hereafter, Paper~II), 
we perform a detailed analysis of this problem and 
provide an analytic solution for the perturbations in a form which is suitable 
for comparison with numerical results.  

This test problem is one-dimensional. We consider a linear, standing 
gravitational wave whose amplitude varies 
in the $z$-direction: 
\beqn
  h_{+}(t,z) &=& h_{+0} \sin \! kz \cos \! kt \ , \label{eq:hplust} \\
  h_{\times}(t,z) &=& h_{\times 0} \sin \! kz \cos \! kt \ ,
\label{eq:hcrosst}
\eeqn
where $k$ is the wave number, and $h_{+0}$ and 
$h_{\times 0}$ are constants. 
We assume that at $t=0$, the magnetized fluid is unperturbed: 
\beqn
  P(0,z) = P_0\ , \ \ \ \ \ \ \ \ \ \ \rho_0(0,z) = \rho_0 \ , 
\label{eq:MHDinit1} \\
  v^i(0,z) = 0\ , \ \ \ \ \ \ \ \ \ \ B^i(0,z) = B_0^i \ . 
\label{eq:MHDinit2}
\eeqn
Subsequently, the gravitational wave excites the MHD modes of the fluid.
As discussed in Paper~II, the gravitational wave is unaffected by the fluid
to linear order, and the metric perturbation, $h_{\mu \nu}(t,z)$, in 
the transverse-traceless (TT) gauge can be calculated from 
Eqs.~(\ref{eq:hplust}) 
and (\ref{eq:hcrosst}). The perturbations in pressure $\delta P(t,z)$, 
velocity $\delta v^i(t,z)$, and magnetic field $\delta B^i(t,z)$ can 
be computed analytically as shown in Paper~II. Our analytic solutions are 
valid as long as we are in the linear regime in which the following three 
inequalities hold (see Paper~II): 
\beqn
 |h^{\mu\nu}| \sim h_0 & \ll & 1 \ , \label{gwmhd_ineq1} \\ 
\frac{|T^{\mu\nu}|}{|h^{\mu\nu}|}  \sim   \frac{\cal E}{h_0} & \ll & k^2 \ , \label{gwmhd_ineq2} \\  
t  \ll   1/\sqrt{|T_{\mu \nu}|} & \sim & 1/\sqrt{\cal E} , \label{gwmhd_ineq3}
\eeqn
where $h_0=\sqrt{h_{+0}^2 + h_{\times 0}^2}$ and 
${\cal E} = \rho_0 (1+\epsilon_0) + P_0 + b_0^2$.

The analytic solution is a superposition of the three eigenmodes of the 
homogeneous system (the Alfv\'en,
slow magnetosonic, and fast magnetosonic waves) and a particular solution which
oscillates at the frequency of the gravitational wave. The induced 
Alfv\'en wave obeys the dispersion relation
\beq
  \omega^2 = \omega_A^2 \equiv
(\ve{k}\cdot \ve{v}_A)^2 \ ,
\eeq
where $\ve{k}=k\hat{\ve{z}}$ is the wave vector associated with 
the standing gravitational wave, and 
$\ve{v}_A = \ve{B}_0/\sqrt{4\pi{\cal E}}$ is
the Alfv\'en velocity. This mode gives rise to a velocity perturbation
\beq
 \delta {\ve{v}} \propto \tilde{\ve{u}}_A \equiv
\ve{k} \times \ve{v}_A .
\eeq
The frequencies of the induced slow and fast magnetosonic modes, 
$\omega_{m1}$ and $\omega_{m2}$,
are found by solving the following dispersion relation for $\omega^2$:
\beq
  \omega^4 - [ k^2 c_m^2 + c_s^2 (\ve{k} \cdot
\ve{v}_A)^2 ] \omega^2
 + k^2 c_s^2 (\ve{k} \cdot \ve{v}_A)^2 = 0 \ ,
\eeq
where $c_m^2 = v_A^2 + c_s^2(1-v_A^2)$.  For the corresponding
eigenvectors, one has:
\beq
 \delta \ve{v} \propto \tilde{\ve{u}}_{mi}
\equiv \ve{v}_A 
+ \frac{\omega_{mi}^2 (1-v_A^2)}{(\omega_{mi}^2-k^2)(\ve{k} \cdot \ve{v}_A)} \ve{k}
 \ \ i=1,2 \ .
\label{eigen:magneto}
\eeq
Note that $\tilde{\ve{u}}_A$ is orthogonal to $\tilde{\ve{u}}_{m1}$ and 
$\tilde{\ve{u}}_{m2}$, but $\tilde{\ve{u}}_{m1}$ and $\tilde{\ve{u}}_{m2}$ 
are not, in general, orthogonal to each other. 

The setup for our code test is as follows. Our computational 
domain is $z \in (-1,1)$. We choose $k = 2\pi$ so that 
our computational domain covers two wavelengths of the gravitational wave. 
Our standard resolution is $\Delta z = 0.01$ (200 grid points). 
At time $t=0$, we assign the 
metric $g_{\mu \nu}(0,z) = \eta_{\mu \nu} + h_{\mu \nu}(0,z)$, where 
$\eta_{\mu \nu} = \rm{diag}(-1,1,1,1)$ is the Minkowski metric, and
the nonzero components of $h_{\mu \nu}(0,z)$ are:
\beqn
 h_{xx}(0,z) &=& -h_{yy}(0,z) = h_+(0,z) \ , \\
 h_{xy}(0,z) &=& h_{yx}(0,z) = h_{\times}(0,z) \ .
\eeqn
We choose geodesic slicing and zero shift ($\alpha=1$, $\beta^i=0$) as our gauge 
conditions. Hence we set the initial extrinsic curvature to zero 
($K_{ij}(0,z)=0$) in accord with our gauge choices and 
Eqs.~(\ref{eq:hplust}), (\ref{eq:hcrosst}), and (\ref{Kij}). We also set the 
MHD variables at $t=0$ according to Eqs.~(\ref{eq:MHDinit1}) and 
(\ref{eq:MHDinit2}). We choose the adiabatic index $\Gamma = 4/3$ 
in all of our simulations in this section. 
Periodic boundary conditions on both matter and
gravitational field quantities are enforced at the upper and lower boundaries
in $z$.
We expect that the metric, as well as 
the MHD quantities, in our full GRMHD simulations should agree 
with the analytic solutions given in Paper~II to linear order 
as long as the 
inequalities in Eqs.~(\ref{gwmhd_ineq1})--(\ref{gwmhd_ineq3}) are satisfied. 

\subsubsection{A General Example}
\label{gwmhd_general}

\begin{figure}
\includegraphics[width=8cm]{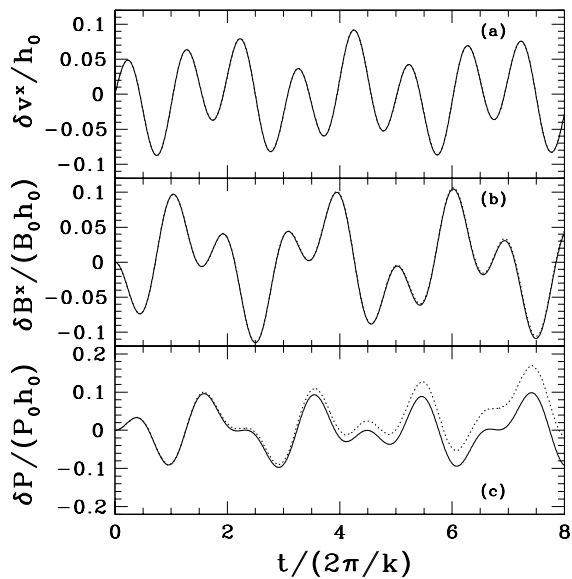}
\vskip 0.5cm
\caption{Analytic and numerical solutions for the perturbations of a 
magnetized fluid due to the presence of a gravitational wave (see
Section~\ref{gwmhd_general}).  The thick solid and thin dotted lines 
represent, respectively, the analytic and numerical solutions, though the 
two lines are not readily distinguishable in plots (a) and (b).  All quantities
are evaluated at $z = 1/8$ and are normalized as indicated.  
Time is normalized by the gravitational wave period. }
\label{gwmhd_comp1}
\end{figure}

We first consider a general case in which all three of the MHD modes
are excited.  We take the following initial data:
\beqn
& & \rho_0 = 2.78\times 10^{-9} \ , \ \ \  P_0 = 1.29 \times 10^{-9} \ ,  \nonumber \\
& & B_0^i = (1.09, 8.26, 14.4) \times 10^{-5} \ ,  \nonumber \\ 
& & h_{+0} = h_{\times 0} = 1.18 \times 10^{-4} \ .
\label{gwdata1}
\eeqn
Figure~\ref{gwmhd_comp1} gives a comparison of the analytic and numerical solutions 
for three selected perturbed variables.  The perturbations are plotted 
with respect to time for a chosen location on the grid ($z=1/8$).  Good agreement is 
shown between the numerical and analytic values for 
many periods of the gravitational wave. We also find very good 
agreement for the metric quantities
$g_{xx}$ and $g_{xy}$ in our simulation and the analytic values 
calculated from Eqs.~(\ref{eq:hplust}) and (\ref{eq:hcrosst}). 
The pressure perturbation, however, 
differs from the analytic solution by a slight secular drift.  (In fact, all 
variables eventually exhibit a drift away from the analytic solution, but the 
drift is first noticeable in the case of the pressure.)  

\vskip 1cm
\vskip 1cm
\begin{figure}
\includegraphics[width=8cm]{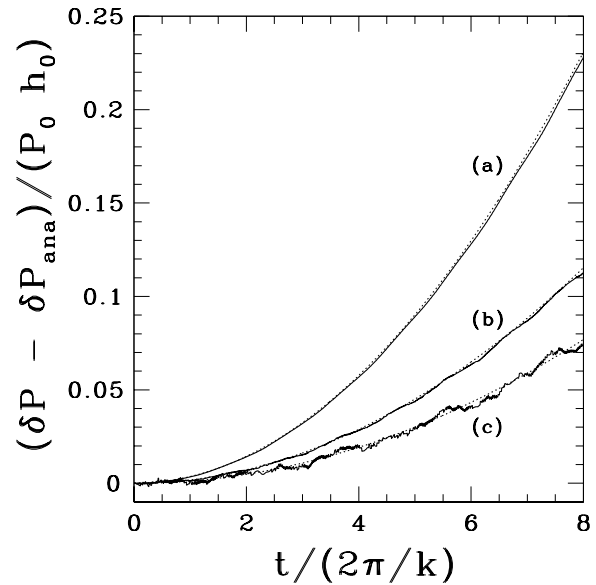}
\vskip 0.5cm
\caption{Relative error in the pressure perturbation (evaluated 
at $z=1/8$) for three different initial data sets.  
The solid lines give the numerical results while the dashed
lines give the fits from Eq.~(\ref{errorpredict}).  The three 
initial data sets are derived from the standard case in Eq.~(\ref{gwdata1}) by taking 
$ \{ \rho_0,\, P_0 \}^{\rm new} = \xi \{ \rho_0,\, P_0 \}^{\rm old}$, 
$(B^i_0)^{\rm new} = \sqrt{\xi} (B^i_0)^{\rm old}$, 
and $\{ h_{+0},\, h_{\times 0} \}^{\rm new} = \zeta \{ h_{+0},\, h_{\times 0} \}^{\rm old} $,
where $\xi$ and $\zeta$ are constants and ``old'' refers to the values
in Eq.~(\ref{gwdata1}).  (Effectively, $T_{\mu\nu}^{\rm new} = \xi T_{\mu\nu}^{\rm old}$ 
and $h_0^{\rm new} = \zeta h_0^{\rm old}$.)  The regime of validity for the analytic solution
is approached for small $h_0$ and $|T_{\mu\nu}|/h_0$ (see Eqs.~(\ref{gwmhd_ineq1}) and (\ref{gwmhd_ineq2})), 
or equivalently, for decreasing $\zeta$ and $\xi/\zeta$.    For the curves shown, these values are:
(a)~$\zeta = 3$, $\xi/\zeta = 3$, (b)~$\zeta = 2$, $\xi/\zeta = 2$, and (c)~$\zeta = 1$, $\xi/\zeta = 1$. 
Moving from (a) to (c), we find that the relative error 
decreases as expected and that the errors are well fit by Eq.~(\ref{errorpredict}). 
Note that the normalization differs in the three cases to reflect 
the differing values of $P_0$ and $h_0$.  }

\label{errorfig}
\end{figure}

\vskip 1cm
\begin{figure}
\includegraphics[width=8cm]{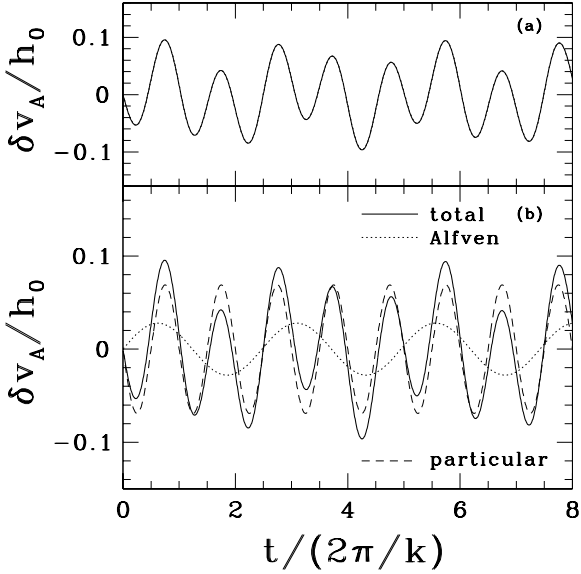}
\vskip 0.5cm
\caption{Velocity projected along the direction of the Alfv\'en mode eigenvector
($\delta v_A \equiv \delta\ve{v}\cdot \tilde{\ve{u}}_A/|\tilde{\ve{u}}_A|$), for
the case discussed in Section~\ref{gwmhd_general} (the general case).
(a) Analytic and numerical plots of the projected velocity. (The curves cannot be
distinguished on this scale.)  (b) Contributions to the analytic solution for 
$\delta v_A /h_0$.  The particular solution (dashed) and Alfv\'en wave (dotted) 
contributions added together give the total perturbation (solid line). 
All quantities are evaluated at $z=1/8$.}
\label{gwmhd_dva}
\end{figure}

\vskip 1cm
\begin{figure}
\includegraphics[width=8cm]{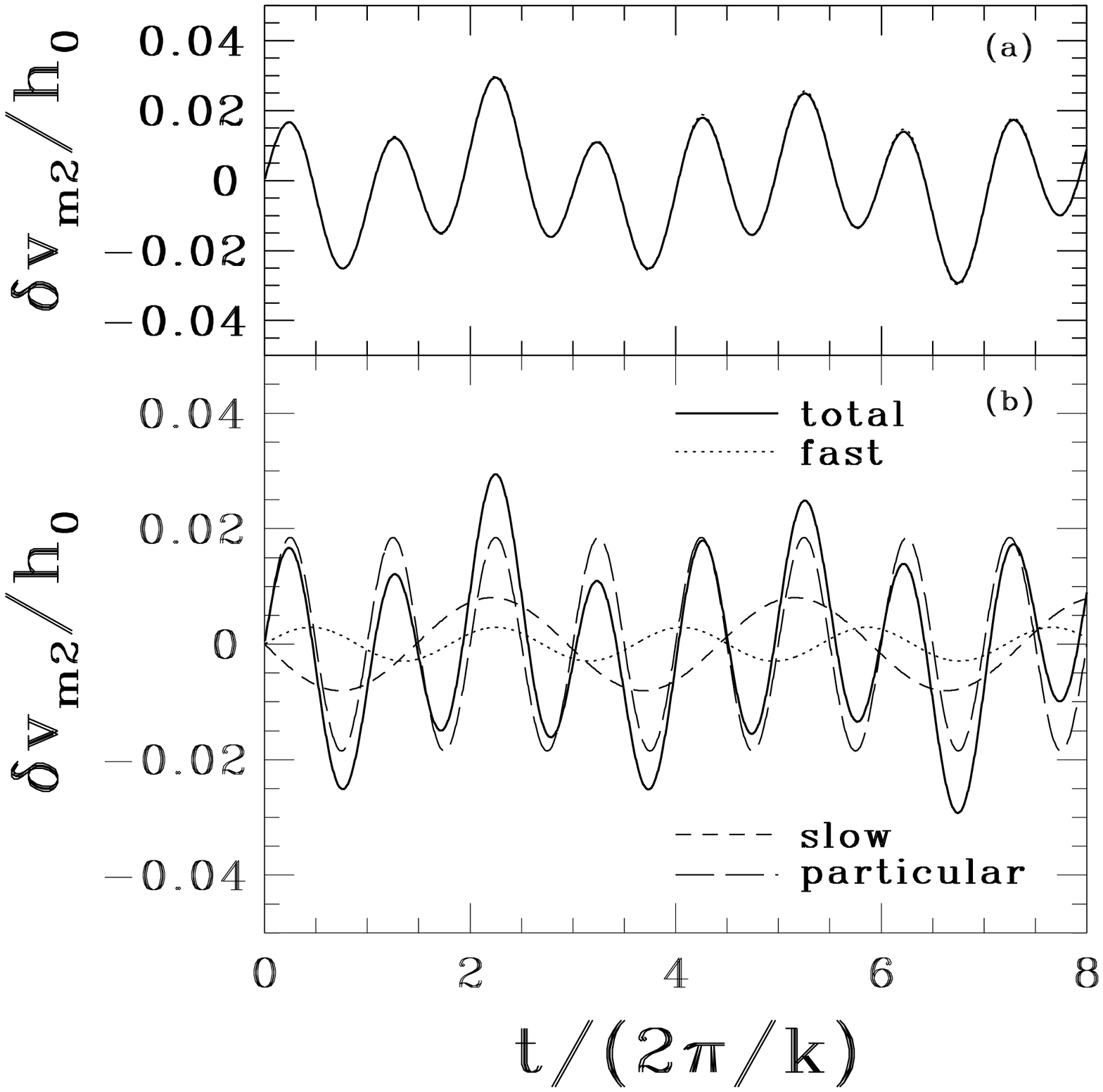}
\vskip 0.5cm
\caption{Velocity projected along the direction of the slow magnetosonic mode 
eigenvector
($\delta v_{m2} \equiv \delta\ve{v}\cdot \tilde{\ve{u}}_{m2}/|\tilde{\ve{u}}_{m2}|$), for
the same case as in Figure~\ref{gwmhd_dva}.
(a) Analytic and numerical plots of the projected velocity. (b) Contributions to 
the analytic solution for 
$\delta v_{m2} /h_0$.  The fast magnetosonic (dotted), slow magnetosonic (short
dashed), and particular solution (long
dashed) contributions added together give the total perturbation (solid line). 
All quantities are evaluated at $z=1/8$.}
\label{gwmhd_dvm2}
\end{figure}

This secular drift is {\em not} due to numerical error, but rather is an effect of
the nonlinear terms which are neglected in our analytic solution.  To show
that it is not a numerical error, we performed simulations at resolutions of
50, 100, and 200 grid points, and we found convergence to second order
to a solution with nonzero drift.
Since the discrepancy is due to nonlinear terms, choosing smaller (larger) 
initial mass-energy density and smaller (larger) gravitational wave strength leads
to a smaller (larger) discrepancy with the analytic solution.  In particular,
the size of the discrepancy is controlled by the degree to which the 
conditions in Eqs.~(\ref{gwmhd_ineq1})--(\ref{gwmhd_ineq3}) are satisfied.
By evolving a range of initial data sets in which we independently varied $h_0$ 
and ${\cal E}/h_0$, as well as initial data sets with $h_0 = 0$
(no gravity wave), we found that the numerical solution for the  pressure perturbation
($\delta P$) is always well fit by the relation 
\beq
\delta P  = \left\{ \begin{array}{ll} \delta P_{\rm ana} + 
h_0 P_0(c_1 h_0 + c_2 {\cal E}/h_0) t^2  & (h_0 \neq 0) \\
c_2 P_0 {\cal E} t^2 & (h_0 = 0) \ , \end{array} \right.
\label{errorpredict}
\eeq
where $t$ is the coordinate time, $\delta P_{\rm ana}$ is the analytic solution for the 
pressure perturbation given in Paper~II, and $c_1$ and $c_2$ are 
constants (see Fig.~\ref{errorfig}).  (Note
that the coefficient $h_0 P_0$ is simply the typical scale of the pressure 
perturbation.)  The term 
proportional to $c_1$ corresponds to nonlinear effects of the gravitational 
wave on the fluid, while the term proportional to $c_2$ is related to the self-gravity 
of the fluid~\cite{footnote:hcon}.  Neither of these effects are accounted for in our analytic solution
[see Eqs.~(\ref{gwmhd_ineq1})--(\ref{gwmhd_ineq3})].  
Thus, the disagreement of the numerical and analytic 
results may be reduced by choosing initial data with smaller $h_0$ and ${\cal E}/h_0$.  

To extract the various MHD modes in this test, 
we have performed two projections of the velocity.  Figure~\ref{gwmhd_dva}a shows 
the numerical and analytic values of the projection along $\tilde{\ve{u}}_A$ 
(again evaluated at $z=1/8$).  Because we have
projected the velocity along the direction of the Alfv\'en mode eigenvector, which
is orthogonal to the fast and slow mode eigenvectors, we only pick up contributions
from the Alfv\'en wave and the particular solution (see Paper~II for the analytic 
expression of the particular solution).  The lower panel shows these 
individual contributions along with the total.  One can see that both the Alfv\'en
and particular components contribute significantly to the velocity perturbation
in this direction.  Next, Fig.~\ref{gwmhd_dvm2} shows the projection of the 
velocity along the direction of the slow mode eigenvector.  This time, there are
contributions from the slow and fast modes in addition to the particular solution,
and one again sees that all modes are contributing strongly.  (The slow and fast
modes are both present in this velocity projection since 
$\tilde{\ve{u}}_{m1}$ and $\tilde{\ve{u}}_{m2}$ are not orthogonal.) 
Figures~\ref{gwmhd_dva} and~\ref{gwmhd_dvm2}, taken together, show that our code
correctly manifests all three
MHD waves, in addition to the particular solution contribution.

\begin{table*}
\caption{Evaluation of Methods}
\begin{center}
\tabcolsep 0.15in
\begin{tabular}{c  c c c}
\hline
\hline
\vspace{0.03in}
Method  &  \multicolumn{3}{c}{Characterization} \\
  & Equilibrium Stars & Shocks & Alfv\'en Waves  \\
\hline\hline
\small{HLL+PPM/HLL+CT}        & \small{performs well}           & \small{performs well}  & \small{performs well} \\
\small{HLL+CENO/HLL+CT}       & \small{central density drifts}  & \small{performs well}  & \small{performs well} \\
\small{HLL+MC/HLL+CT }        & \small{central density drifts}  & \small{performs well}  & \small{performs well} \\
\small{HLL+MC/VP}             & \small{central density drifts}  & \small{unacceptable oscillations}  & \small{acceptable} \\
\small{non-conservative/CT}   & \small{performs well}  & \small{problems for high $b^2/\rho_0$}  & \small{problems for high $b^2/\rho_0$} \\
\hline
\hline
\end{tabular}
\end{center}
\begin{minipage}{14cm}
\raggedright
\end{minipage}
\label{tab:summary}
\end{table*}

\subsubsection{Special Cases}
\label{gwmhd_special}
To further test our code, we consider several initial data sets with special
properties.  Starting
with the initial data set given in Eq.~(\ref{gwdata1}), we hold ${\cal E}$ and
$h_0$ constant but change the balance between the plus and cross modes 
so that the magnetosonic modes are not excited.  As explained in Paper~II
[in particular, see Eq.~(93) and surrounding discussion], this occurs when 
$h_{+0}$ and $h_{\times 0}$ satisfy the equation
\beq
  [(B_0^y)^2 - (B_0^x)^2] h_{+0} - 2B_0^x B_0^y h_{\times 0} = 0 \ .
\label{cond:SGWz0}
\eeq
Thus, we obtain the new initial data set:
\beqn
& & \rho_0 = 2.78\times 10^{-9} \ ,  \ \ \  P_0 = 1.29 \times 10^{-9} \ , \nonumber \\
& & B_0^i = (1.09, 8.26, 14.4)\times 10^{-5} \ ,  \nonumber \\ 
& & h_{+0} =  4.31 \times 10^{-5} \ \ \  h_{\times 0} = 1.61 \times 10^{-4} \ .
\label{gwdata2}
\eeqn
In the analytic 
solution for this case, the pressure perturbation vanishes identically.  
In accord with this, our numerical solution for the pressure perturbation
shows no oscillations, though the slight secular drift is still present.  Similarly, 
the projection of the velocity along the slow mode eigenvector vanishes up 
to some small-amplitude noise, but the projection along the Alfv\'en mode 
eigenvector does
not vanish and the analytic and numerical solutions agree very well.  Thus,
by changing only the relative proportion of the plus and cross modes, we
have arrived at a very different physical outcome from the more general case 
described in Section~\ref{gwmhd_general}, and our code again correctly 
identifies the modes which are present.

The analytic solutions in Paper~II also indicate that the gravitational 
wave has no effect on the fluid if (1) $B_0^x=B_0^y=0$, or (2) $B_0^z=0$ 
and $B_0^x$ and $B^y_0$ satisfy Eq.~(\ref{cond:SGWz0}). We have performed 
simulations in these two special cases and found that our numerical 
solutions for the perturbations contain only small amplitude noise or 
the secular drift due to nonlinear effects, as expected.

\section{Conclusions}
We have developed the first code which is able to evolve the full coupled
Einstein-Maxwell-MHD equations in 3+1 dimensions without 
approximation. Our code is able to model the behavior of magnetized,
perfectly-conducting fluids in dynamical
spacetimes. We have confirmed the ability of this code to accurately simulate
unmagnetized hydrodynamic stars, MHD shocks, Alfv\'en waves, 
magnetized accretion onto a black hole, and the excitation of MHD modes 
in a magnetized fluid driven by
gravitational waves. We have performed 1, 2, and 3 dimensional tests.

We have tested several different integration schemes. 
In Table~\ref{tab:summary}, we evaluate the behavior of each method under
various tests.  The first row describes the results obtained by using our
``standard'' code (listed as HLL+PPM/HLL+CT), in which
the fluid equations are evolved using HLL fluxes and PPM reconstruction, while
the magnetic induction equation is evolved using HLL fluxes interpolated 
to preserve the constraints (CT). The next two rows describe schemes which 
differ from HLL+PPM/HLL+CT
only in the reconstruction method used with the hydrodynamic variables
($\rho_{\star}$, $\tilde\tau$, and $\tilde S_i$):
HLL+CENO/HLL+CT uses CENO reconstruction, and HLL+MC/HLL+CT uses MC reconstruction. 
We have also experimented with more significant changes.  HLL+MC/VP evolves the
fluid variables with HLL fluxes and MC reconstruction (like HLL+MC/HLL+CT), 
but the induction equation is solved by evolving the vector potential $A^i$.  
In this method, the
magnetic field is automatically divergence-free, because it is calculated 
as the curl of the vector potential: $B^i = n_{\mu} \epsilon^{\mu ijk} 
A_{k,i}$.
Finally, we test a nonconservative MHD scheme,
non-conservative/CT, which is a straightforward extension of our previous 
hydrodynamics code~\cite{dmsb03} to MHD.  Non-conservative/CT uses flux-CT
to maintain constrained transport. 
It differs from the non-conservative MHD code of~\cite{dVh03} in that 
our grid is not staggered, our energy variable is different,
and our time integration is done differently.  From the table, we
draw several conclusions.
 (1) PPM is the best of the reconstruction methods considered here.  Using
     PPM reconstruction, we are able to achieve accurate evolutions, even
     with a simple (HLL) Rieman solver.
 (2) Evolving the magnetic field directly with constrained transport gives better
     results than evolving a vector potential.  The vector potential method performs
     especially poorly in the presence of shocks.
 (3) Our nonconservative method works well for problems involving only weak magnetic
     fields, but it becomes unstable for problems in which the magnetic energy density
     significantly exceeds the gas energy density.  It is, therefore, not suitable
     for some problems.  (Note, however, that better
     nonconservative MHD codes have been developed~\cite{dVh03} which can accommodate
     larger magnetic fields.)

Our MHD code has limitations similar to those of other MHD codes in the literature.  In
particular, accurate evolution is difficult when $b^2 \gg \rho_0$.  This could potentially
cause problems in the low-density regions in some applications.  However, the experience
of other numerical MHD groups suggests that these difficulties are surmountable.

Having passed the tests described above, we will next apply our code
to the study of self-gravitating magnetized fluids in astrophysical problems. 
In particular, we plan to model the braking of differential rotation described
in several applications in the introduction.  We also plan to simulate gravitational
collapse in order to explore the behavior and influence of magnetic fields on
supernovae and collapsars.

\acknowledgments
It is a pleasure to thank C.~Gammie, J.~McKinney, S.~Noble, and J.~Hawley
for useful suggestions and discussions. We also thank S.~Komissarov for 
geneously providing us with codes to compute the exact solutions to 
the problems in Section~\ref{kom_tests}
to which our numerical results were compared.  Some of our calculations
were performed at the National Center for Supercomputing Applications at the
University of Illinois at Urbana-Champaign (UIUC). 
This paper was supported in part by NSF Grants PHY-0205155 and PHY-0345151,
and NASA Grants NNG04GK54G and NNG046N90H.

\appendix

\section{Implementation of PPM reconstruction}
\label{ppm_appendix}

The piecewise parabolic method (PPM) is an algorithm used to
construct the values of a primitive variable $p$ to the left
and right of each zone interface ($p^L_{i+1/2} = p_{i+1/2-\epsilon}$
and $p^R_{i+1/2} = p_{i+1/2+\epsilon}$).  It consists of several steps. 
First, one interpolates to $p_{i+1/2}$ according to
\beqn
\label{ppm_interp}
p^L_{i+1/2} &=& p^R_{i+1/2} = p_{i+1/2} \\ 
&=& \nonumber p_i + {1\over 2}(p_{i+1}-p_i)
            + {1\over 8}\Delta x(\nabla p_i - \nabla p_{i+1})\ ,
\eeqn
where $\nabla p$ is the MC slope-limited gradient of $p$ 
(see Eq.~(\ref{eq:mc})).  Note that the factor ${1\over 8}$ on the
last term differs from the ${1\over 6}$ sometimes appearing in the
literature.  We find more accurate results with the ${1\over 8}$ in
Eq.~(\ref{ppm_interp}). 
Next, $p^L_{i+1/2}$ and $p^R_{i+1/2}$ are adjusted using ``steepening'', 
``flattening'', and ``monotonizing'' algorithms, which are intended to
stabilize the evolution and sharpen shock profiles.  There are several
adjustable parameters in the PPM scheme; we use the values recommended
in ~\cite{PPM}.

As originally proposed, PPM reconstruction reduces to first-order accuracy 
at extrema of $p$.  This is due to a ``monotonization'' step of the
PPM algorithm which removes local extrema in the interpolation function
in order to suppress unphysical oscillations near shocks.  Near a maximum
or minimum of $p$, however, the extremum in the interpolation function
represents the true behavior in $p$.  Therefore, we have followed~\cite{c88}
in distinguishing between local extrema caused by numerical oscillations
and physical extrema.  When the first
derivative $\Delta x^{-1}(p_{i+1} - p_i)$ changes sign, but the
second derivative does not change sign over two grid cells in either
direction, the extremum is regarded as a physical maximum or minimum and
the standard monotonization routine is not applied.  Third-order accuracy
may still be sacrificed at these points by other steps of the PPM algorithm,
but we have found that this modification significantly improves accuracy
for the evolution of stars.  When interpolating $\rho_0$, we also
turn off the monotonization when $\rho_0$ is within 15\% of its
maximum value on the grid.  We do this to make sure we do not lose
accuracy at the centers of our stars, at which the density is usually
a global maximum.

Because the $p^L_{i+1/2}$ and $p^R_{i+1/2}$ are equal (and hence 
$u_R=u_L$) at most cell
interfaces, PPM usually picks up no dissipation from the $(u_R - u_L)$
term in the HLL flux formula (Eq.~(\ref{eq:hll})).  Usually, this is a
good thing.  For cases where some extra dissipation is desirable, such
as in the relativistic accretion test, we add a small Kreiss-Oliger
dissipation.  This takes the form
\begin{equation}
\partial_t u = \cdots
- C_{\rm ko}{(\Delta X\Delta Y\Delta Z)^{4/3}\over 16\Delta T} 
\nabla^2_f (\nabla^2_f u)\ ,
\end{equation}
where $\nabla^2_f$ is the flat-space Laplacian.  We have found good results
with $C_{\rm ko} \sim 0.1$.

\section{Setup of initial data for the nonlinear Alfv\'en wave}
\label{app:alfven}

One exact solution of the MHD equations is an Alfv\'{e}n wave traveling
in Minkowski spacetime.  The solution was derived by Komissarov~\cite{k97}. 
Here we summarize the results.  Suppose that the position of
the wavefront is given by the phase function
\beq
\Phi(x^{\alpha}) = 0\ .
\eeq
In the Lorentz frame of interest, let $\mu$ be the wave speed and
${\bf n}$ be the unit three-vector in the direction of propagation
of the wave front. Then, with the appropriate scaling of $\Phi$, 
we define 
\beq
\Phi_{\alpha} \equiv \Phi_{,\alpha} = (-\mu,{\bf n})\ . 
\eeq
Note that $\Phi_{\alpha}$ is proportional to the 1-form $k_{\alpha} = (-\omega,\ve{k})$,
which is dual to the propagation 4-vector $k^{\alpha}$. 
We define the following scalars
\beqn
a &=& u^{\alpha}\Phi_{\alpha} \cr \cr
{\cal B} &=& b^{\alpha}\Phi_{\alpha} \cr \cr
{\cal E} &=& \rho_0h + b^2 \label{eq:abe}
\eeqn
The wave speed $\mu$ can be computed from the equation (see, e.g., 
Eq.~(23) of~\cite{k99})
\beq
{\cal E} a^2 - {\cal B}^2 = 0\ . 
\label{eq:wave-speed}
\eeq
Consider two arbitrary points $\ve{r}_+$ and $\ve{r}_-$ connected by
a simple Alfv\'{e}n wave, then (see~\cite{k97})
\beqn
[P] &=& [\rho_0] = [b^2] = [\mu] = 0\ , \cr \cr
[a] &=& [{\cal B}] = 0\ , \cr \cr
[u^{\alpha}] &=& {a\over {\cal B}}[b^{\alpha}]\ ,
\label{eq:alfven_sol}
\eeqn
where $[f] = f({\bf r}_+) - f({\bf r}_-)$. Below, we
will write $f = f({\bf r}_+)$ and $f_- = f({\bf r}_-)$. 

Thus, the Alfv\'{e}n wave perturbation at ${\bf r}_+$ is
specified by one 4-vector, $[b^{\alpha}]$ or $[u^{\alpha}]$.
This 4-vector has one freely specifiable degree
of freedom (corresponding to the amplitude), the other three being
removed by the following constraints:
\beqn
u^{\alpha}u_{\alpha} &=& -1\ , \cr \cr
u^{\alpha}b_{\alpha} &=& 0\ , \cr \cr
[a] &=& 0\ .
\eeqn
Substituting $u^{\alpha} = u_-^{\alpha} + [u^{\alpha}]$ and
$b^{\alpha} = b_-^{\alpha} + {{\cal B}\over a}[u^{\alpha}]$,
we see that these can be rewritten
\beqn
2u^{\alpha}_-[u_{\alpha}] + [u^{\alpha}][u_{\alpha}] &=& 0\ , \cr \cr
(b^{\alpha}_- - {{\cal B}\over a} u^{\alpha}_-)[u_{\alpha}] &=& 0\ , \cr \cr
\Phi_{\alpha}[u^{\alpha}] &=& 0\ ,
\eeqn
One solution for two states connected by an Alfv\'{e}n wave is given 
by Komissarov~\cite{k97}. 

The properties of nonlinear Alfv\'{e}n waves are easiest to study in the wave
frame ($\mu = 0$). We denote the quantities in this frame by a prime. 
If we choose the $x$-axis to be normal to the wave front
and assume $g_{\mu\nu} = \eta_{\mu\nu}$,
then in this frame $a = {u'}^x$, ${\cal B} = {b'}^x$, and we have
\beqn
[{u'}^x] &=& 0, \qquad [{b'}^x] = 0\ , \cr \cr
\chi &=& {u'}^x/{b'}^x\ , \cr \cr
[{u'}^{\alpha}] &=& \chi [{b'}^{\alpha}]\ .
\eeqn
The divergence condition on ${B'}^i$ requires that $[{B'}^x] = 0$.
Solving the other constraints, one finds that the transverse components 
of ${b'}^i$ lie on the ellipse
\beqn
\label{b_ellipse}
& & a_{11} {b'}_y^2 + (a_{12} + a_{21})b'_y b'_z + a_{22} {b'}_z^2 \cr \cr
& & \quad + (a_{13} + a_{31})b'_y + (a_{23} + a_{32})b'_z + a_{33} = 0\ ,
\eeqn
where
\beqn
a_{11} = 1 - a_y^2\ ,       &\qquad& a_{22} = 1 - a_z^2\ , \cr \cr
a_{33} = -(c^2 + d)\ ,      &\qquad& a_{12} = a_{21} = -a_y a_z\ , \cr \cr
a_{13} = a_{31} = -c a_y\ , &\qquad& a_{23} = a_{32} = -c a_z\ ,
\eeqn
and where
\beqn
a_y &=& {{u'}^y_- - \chi {b'}^y_-\over {u'}^0_- - \chi {b'}^0_-}\ , \qquad
a_z = {{u'}^z_- - \chi {b'}^z_-\over {u'}^0_- - \chi {b'}^0_-}\ , \cr \cr
c &=& {\chi b^2 \over {u'}^0_- - \chi {b'}^0_-}\ , \qquad d = b^2 - {b'}_x^2\ .
\eeqn
The center of this ellipse is at
\beq
(b^y_c, b^z_c) = \left({c\over D}\right)(a_y, a_z)\ ,
\eeq
where
\beq
D = {{b'}_x^2 - b^2 {u'}_x^2\over {B'}_x^2}\ .
\eeq
It is convenient to rewrite the equation of the ellipse~(\ref{b_ellipse}) 
in terms of a free parameter $\theta$ defined by
\beqn
 {b'}^y &=& b^y_c + b_{yz}(\theta) \cos \theta \ , \label{eq:bpy} \\
 {b'}^z &=& b^z_c + b_{yz}(\theta) \sin \theta \ . \label{eq:bpz}
\eeqn
Substituting this into Eq.~(\ref{b_ellipse}), one obtains
\beq
  b_{yz}(\theta) = \sqrt{ \frac{d+c^2/D}{a_{11}\cos^2 \theta + 2 a_{12} 
\sin \theta \cos \theta + a_{22} \sin^2 \theta} } \ .
\label{eq:byz}
\eeq

One can construct an Alfv\'en wave (propagating in $x$-direction) connecting 
a left state and a right state as follows: 
\begin{enumerate}
\item Choose the width $W$ that connects the left and right state. In 
our test, we choose $W=0.5$.
\item Choose $\rho_0$ and $P$, which are constant throughout the wave.
\item Choose $B_-^i$ and $u_-^{\mu}$ on the left side. Without loss of
generality, one may set $B_-^z=0$.
\item Calculate $b_-^{\mu}$ from Eqs.~(\ref{bintermsofB}) and
(\ref{eq:baBa}).  
\item Calculate $\cal B$, $a$, $\cal E$ and the wave speed $\mu$ from 
Eqs.~(\ref{eq:abe}) and (\ref{eq:wave-speed}).
\item Compute ${b'}_-^{\mu}$ and ${u'}_-^{\mu}$ by boosting 
$b_-^{\mu}$ and $u_-^{\mu}$ to the wave frame. Then compute ${B'}_-^i$ from 
${b'}_-^{\mu}$ and ${u'}_-^{\mu}$. 
\item Set ${u'}^x(x)={u'}_-^x$ and ${b'}^x(x)={b'}_-^x$ 
(constant everywhere). 
\item Compute $\chi$, $a_y$, $a_z$, $c$, $d$, $a_{ij}$. 
\item Compute $b^y_c$, $b^z_c$. 
\item Choose $\theta(x)$ consistent 
with $({b'}_-^y,{b'}_-^z)$. In this paper, we choose~\cite{fn1} 
\beq
  \theta(x) = \left \{ \begin{array}{ll} \theta_l  & x \leq -W/2 \\ 
 & \\
   \theta_l + A \sin^2 \left[\frac{\pi(x+W/2)}{2W}\right] &
-W/2 \leq x \leq W/2 \\ & \\
   \theta_l + A & x \geq W/2 \end{array} \right.  \ ,
\label{eq:theta_x}
\eeq
where $A$ is a freely specifiable constant (``amplitude'' of the wave) 
and $\theta_l$ is given by 
\beq
  \theta_l = \tan^{-1} \left( \frac{ {b'}_-^z - b^z_c}{{b'}_-^y 
- b^y_c} \right) \ . \label{eq:theta_l}
\eeq
One can verify
that our choice of $\theta(x)$ gives the correct left state, and the
right state is determined by the value of $A$.
We choose $A=\pi$ in our Alfv\'en wave test in order to compare our
numerical results with those by Komissarov~\cite{k99}. This means that 
the tangential 
component of ${b'}^{\mu}$ is rotated by $\pi$ when going from 
the left state to the right state.
\item Compute ${b'}^y(x)$ and ${b'}^z(x)$ from 
Eqs.~(\ref{eq:bpy})--(\ref{eq:byz}), (\ref{eq:theta_x}) 
and (\ref{eq:theta_l}). 
\item Use $[{u'}^{\alpha}] = \chi [{b'}^{\alpha}]$ to set $({u'}^y,{u'}^z)$ 
as a function of $x$. 
\item Compute ${u'}^0(x)$ and ${b'}^0(x)$ from the relations 
${u'}^{\mu} {u'}_{\mu} =-1$ and ${b'}^{\mu} {u'}_{\mu}=0$.
\item Compute $b^{\mu}(x)$ and $u^{\mu}(x)$ by boosting 
${b'}^{\mu}(x)$ and ${u'}^{\mu}(x)$ back to the original frame.
\item Compute $B^i(x)$ from $u^{\mu}(x)$ and $b^{\mu}(x)$.
\end{enumerate}

We use this recipe to construct the initial data for our nonlinear 
Alfv\'en wave test.


\begin{thebibliography}{67}
\bibitem{footnote:pvar}
  In the MHD case, the primitive variable inversion can also be reduced to a single
  polynomial equation for a $\Gamma$-law EOS~\cite{dZbl03,pvarMHD}. 
  However, this inversion scheme is not as robust as multi-dimensional
  Newton-Raphson solvers~\cite{pvarMHD}.
\bibitem{footnote:ex}
  In order to make the collapse occur at the same time at all resolutions,
  we deplete a small percentage (2\%) of the initial pressure for these runs. 
  This depletion is small enough to not significantly affect the violation of
  the constraint equations.  Pressure depletion was not used for the star \starb\
  run in Fig.~\ref{fig:starstability}; resolution-dependent numerical error is
  sufficient to trigger collapse.
\bibitem{fn:bondi} Note that the initial metric in the Bondi test 
accounts only for the black hole and not the fluid. This introduces 
small initial constraint violations, which slightly affect the metric 
evolution and the rate of growth of the irreducible mass of the black 
hole. We evolved the spacetime
and disk self-consistently in several runs just to show that the
MHD-metric coupling does not introduce numerical instabilities.
\bibitem{footnote:hcon}
We note, however, that the self-gravity of the fluid is not treated
self-consistently in these tests as the initial data do not 
satisfy the Hamiltonian constraint beyond linear order.
\bibitem{fn1} Note that $x$ is the coordinate in the ``laboratory'' frame, 
not in the wave frame. The proper separation between the left and right 
states in the wave frame is longer than $W$ because of length contraction.
\end{thebibliography}
\end{document}